\documentclass[smallextended]{svjour3}       
\smartqed  
\usepackage{graphicx}
\usepackage{mathptmx}      
\usepackage{latexsym}
\usepackage{amsmath}
\usepackage{amssymb}
\usepackage{amsfonts}

\usepackage{color}
\usepackage{fontenc}

\usepackage{txfonts}

\usepackage{natbib}
\usepackage{url}

\newtheorem{observation}{Observation}

\newcommand{\Dom}{\ensuremath{\mathit{Dom}}}

\newcommand{\fr}{\ensuremath{\mathit{fr}}}

\newcommand{\MI}{\ensuremath{\mathit{MI}}}

\newcommand{\zpos}{\ensuremath{\mathit{z_{pos}}}}
\newcommand{\zneg}{\ensuremath{\mathit{z_{neg}}}}
\newcommand{\pbin}{\ensuremath{\mathit{p_{bin}}}}
\newcommand{\pbinI}{\ensuremath{\mathit{p_{bin1}}}}
\newcommand{\pbinII}{\ensuremath{\mathit{p_{bin2}}}}
\newcommand{\pmul}{\ensuremath{\mathit{p_{mul}}}}
\newcommand{\pdouble}{\ensuremath{\mathit{p_{double}}}}

\newcommand{\ctable}{\ensuremath{\mathit{\tau}}}
\newcommand{\obsctable}{\ensuremath{\mathit{\tau_o}}}
\newcommand{\measure}{\ensuremath{\mathit{T}}}
\newcommand{\measureval}{\ensuremath{\mathit{t}}}

\newcommand{\data}{\ensuremath{\mathcal{D}}}
\newcommand{\Model}{\ensuremath{\mathcal{M}}}
\newcommand{\sspace}{\ensuremath{\mathcal{S}}}

\newcommand{\Aisa}{\ensuremath{A{=}a}}

\newcommand{\Bisb}{\ensuremath{B{=}b}}
\newcommand{\Bneqb}{\ensuremath{B{\neq}b}}
 
\newcommand{\Xset}{\ensuremath{\mathbf{X}}}
\newcommand{\Yset}{\ensuremath{\mathbf{Y}}}
\newcommand{\Yprimeset}{\ensuremath{\mathbf{Y'}}}
\newcommand{\Zset}{\ensuremath{\mathbf{Z}}}
\newcommand{\Qset}{\ensuremath{\mathbf{Q}}}
\newcommand{\Rset}{\ensuremath{\mathbf{R}}}

\newcommand{\xbf}{\ensuremath{\mathbf{x}}}
\newcommand{\ybf}{\ensuremath{\mathbf{y}}}
\newcommand{\zbf}{\ensuremath{\mathbf{z}}}
\newcommand{\qbf}{\ensuremath{\mathbf{q}}}
\newcommand{\Xisx}{\ensuremath{\mathbf{X{=}x}}}
\newcommand{\Yisy}{\ensuremath{\mathbf{Y{=}y}}}
\newcommand{\Zisz}{\ensuremath{\mathbf{Z{=}z}}}
\newcommand{\Qisq}{\ensuremath{\mathbf{Q{=}q}}}

\newcommand{\row}{\ensuremath{\mathbf{r}}}

\newcommand{\avgsamp}{\ensuremath{\hat{\mu}}}
\newcommand{\sdsamp}{\ensuremath{\hat{\sigma}}}
\newcommand{\Pest}{\ensuremath{\hat{P}}}

\journalname{}
\begin{document}

\title{A Tutorial on Statistically Sound Pattern Discovery\thanks{This research was supported by the Academy of Finland (decision number 307026).}}


\author{Wilhelmiina H{\"a}m{\"a}l{\"a}inen \and Geoffrey I. Webb}


\institute{W. H{\"a}m{\"a}l{\"a}inen \at
              Aalto University 
              Department of Computer Science, 
              P.O.Box 15400, 
              FI-00076 Aalto, Finland\\
              Tel.: +358-400982787, 
              \email{wilhelmiina.hamalainen@iki.fi}\\
\and
           G.I. Webb \at
              Monash University 
              Faculty of Information Technology, 
              P.O. Box 63, 
              Victoria 3800, Australia\\
              Tel.: +61-399053296, 
              Fax: +61-399055159, 
              \email{geoff.webb@monash.edu}
}

\date{}

\maketitle

\begin{abstract}
\sloppy Statistically sound pattern discovery harnesses the rigour of
statistical hypothesis testing to overcome many of the issues that
have hampered standard data mining approaches to pattern
discovery. Most importantly, application of appropriate statistical
tests allows precise control over the risk of false discoveries --
patterns that are found in the sample data but do not hold in the
wider population from which the sample was drawn. Statistical tests
can also be applied to filter out patterns that are unlikely to be
useful, removing uninformative variations of the key patterns in the
data.  This tutorial introduces the key statistical and data mining
theory and techniques that underpin this fast developing field.  

We concentrate on two general classes of patterns: dependency rules
that express statistical dependencies between condition and consequent
parts and dependency sets that express mutual dependence between set
elements. We clarify alternative interpretations of statistical
dependence and introduce appropriate tests for evaluating statistical
significance of patterns in different situations. We also introduce
special techniques for controlling the likelihood of spurious
discoveries when multitudes of patterns are evaluated.

The paper is aimed at a wide variety of audiences.  It provides the
necessary statistical background and summary of the state-of-the-art
for any data mining researcher or practitioner wishing to enter or
understand statistically sound pattern discovery research or
practice. It can serve as a general introduction to the field of
statistically sound pattern discovery for any reader with a general
background in data sciences.
\end{abstract}



\section{Introduction}
\label{sec:intro}

Pattern discovery is a core technique of data mining that aims at finding all patterns of a specific type that satisfy certain constraints in the data \citep{agrawalass,cooley1997web,rigoutsos1998combinatorial,kim2010human}. Common pattern types include frequent or correlated sets of variables, association and correlation rules, frequent subgraphs, subsequencies, and temporal patterns. Traditional pattern discovery has emphasized efficient search algorithms and computationally well-behaving constraints and pattern types, like frequent pattern mining \citep{AggarwalHanbook}, and less attention has been
paid to the statistical validity of patterns. This has also restricted the use of pattern discovery in many applied fields, like bioinformatics, where one would like to find certain types of patterns without risking costly false or suboptimal discoveries. As a result, there has emerged a new trend towards 
{\em statistically sound pattern discovery} with strong emphasis on statistical validity. In statistically sound pattern discovery, the first priority is to find genuine patterns that are likely to reflect properties of the underlying population and hold also in future data. Often the pattern types are also different, because they have been dictated by  
the needs of application fields rather than computational properties.

\begin{figure*}
\begin{center}
\includegraphics[width=0.9\textwidth]{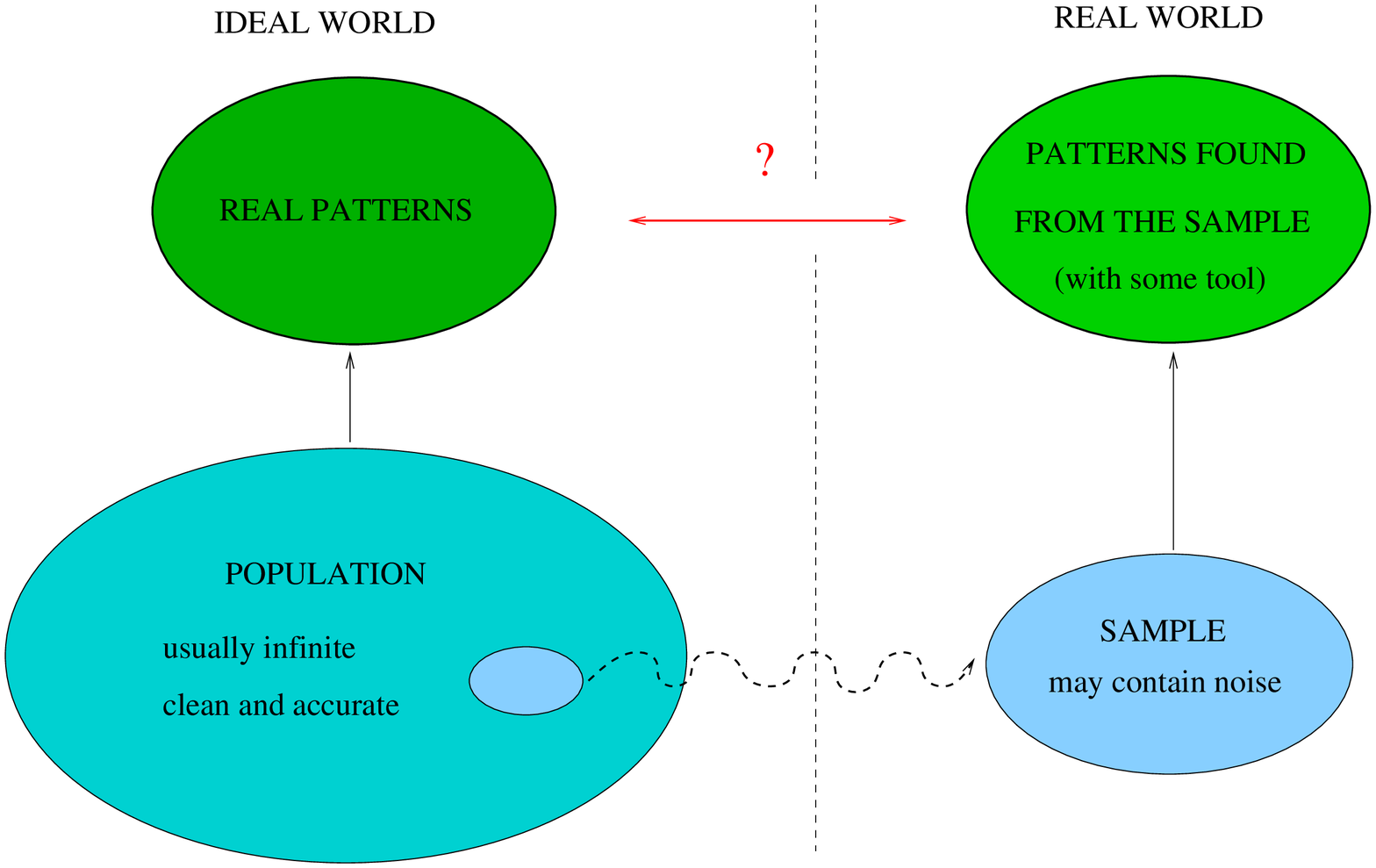}
\end{center}
\caption{An illustration of the problem of finding true patterns from sample data.}
\label{fig:platon}
\end{figure*}%

The problem of statistically sound pattern discovery is illustrated in Fig.~\ref{fig:platon}.  
Usually, the analyst has a sample of data drawn from some population
of interest.  This sample is typically only a very small proportion of
the total population of interest and may contain noise.  The pattern
discovery tool is applied to this sample, finding some set of
patterns.  It is unrealistic to expect this set of discovered patterns
to directly match the ideal patterns that would be found by direct
analysis of the real population rather than a sample thereof.  Indeed,
it is clear that in at least some cases, the application of naive
techniques results in the majority of patterns found being only 
spurious artifacts.  
An extreme example of this problem arises with the popular minimum support and minimum confidence technique \citep{agrawalass} when applied to the well-known Covtype benchmark dataset from the UCI repository \citep{MLrep}. The minimum support  and minimum confidence technique seeks to find the frequent positive dependencies in data using thresholds for minimum frequency ('support') and precision ('confidence').  For the Covtype dataset, the top 197,183,686 rules found by minimum support and minimum confidence are in fact negative dependencies \citep{webb06}.
This gives rise to the suggestion that
the oft cited problem of pattern discovery finding unmanageably large
numbers of patterns is largely due to standard techniques returning
results that are dominated by spurious patterns \citep{Webb11}.

There is a rapidly growing body of pattern discovery techniques being
developed in the data science community that utilize statistics to
control the risk of such spurious discoveries.  This tutorial paper
grew out of tutorials presented at ECML PKDD 2013
\citep{sspdtutorial2013} and KDD-14
\citep{hamalainen2014statistically}. It introduces the relevant
statistical theory and key techniques for statistically sound pattern
discovery. We concentrate on pattern types that express statistical
dependencies between categorical attributes, such as {\em dependency
  rules} (dependencies between condition and consequent parts) and
{\em dependency sets} (mutual dependencies between set elements).
The same techniques of testing statistical significance of dependence also apply to situations where one would like to test dependencies in other types of patterns, like dependencies between subgraphs and classes or between frequent episodes.

To keep the scope manageable, we do not describe actual search algorithms but merely 
the statistical techniques that are employed during the search. 
We aim at a generic presentation that is not bound to any specific
search method, pattern type or school of statistics. Instead, we try
to clarify alternative interpretations of statistical dependence and
the underlying assumptions on the origins of data, because they often
lead to different statistical methods and also different patterns to
be selected. We describe the preconditions, limitations, strengths,
and shortcomings of different approaches to help the reader to select
a suitable method for the problem at hand.  However, we do not make
any absolute recommendations, as there is no one correct way to test
statistical significance or reliability of patterns. Rather, the
appropriate choice is always problem-dependent.

The paper is aimed at a wide variety of audiences. The main goal is to
offer a general introduction to the field of statistically sound
pattern discovery for any reader with a general background in data
sciences. Knowledge on the main principles, important concerns, and
alternative techniques is especially useful for practical data miners
(how to improve the quality or test the reliability of discovered
patterns) and algorithm designers (how to target the search into the
most reliable patterns). Another goal is to introduce possibilities of
pattern discovery to researchers in other fields, like bioscientists,
for whom statistical significance of findings is the main concern and
who would like to find new useful information from large data
masses. As a prerequisite, we assume knowledge of the basic concepts
of probability theory. The paper provides the necessary statistical
background and summary of the state of the art, but a knowledgeable
reader may well skip preliminaries (Sections 2.2-2.3) and the overview
of multiple hypothesis testing (Section 6.1).

In the paper we have tried to use terminology that is consistent with
statistics for two reasons. First, knowing statistical terms makes it
easier to consult external sources, like textbooks in statistics, for
further knowledge. Second, common terminology should make the paper
more readable to wider audience, like reseachers from applied science
who would like to extend their repertoire of statistical analysis with
pattern discovery techniques. To achieve this goal, we have avoided
some special terms originated in pattern discovery that have another
meaning in statistics or may become easily confused in this context
(see Appendix).

The rest of the paper is organized as follows.
In Section~\ref{sec:prel} we give definitions of various types of
statistical dependence and introduce the main principles and
approaches of statistical significance testing.
In Section~\ref{sec:deprules} we investigate how statistical significance of 
dependency rules is evaluated under different assumptions. Especially, we
contrast two alternative interpretations of dependency rules that are
called {\em variable-based} and {\em value-based interpretations} and introduce
appropriate tests for different situations. In Section~\ref{sec:redsig} we discuss
how to evaluate statistical significance of the improvement of one
rule over another one. In Section~\ref{sec:sets} we survey the key techniques that have 
been developed for finding different types of statistically significant dependency sets.
In Section~\ref{sec:multtest} we discuss the problem of multiple hypothesis testing. 
We describe the main principles and popular correction
methods and then introduce some special techniques for increasing power
in the pattern discovery context. Finally, in Section~\ref{sec:concl}  
we summarize the main points and present conclusions. 

\section{Preliminaries}
\label{sec:prel}

In this paper, we consider patterns that express statistical
dependence. Dependence is usually defined in the negative, as absence of 
independence.  Therefore, we begin by defining different types of
statistical independence that are needed in defining dependence
patterns and their relationships, like improvement of one pattern over
another one. After that, we give an overview of the main principles and
approaches of statistical significance testing. These approaches are
applicable to virtually any pattern type, but we focus on how they are
used in independence testing. In the subsequent sections, we will
describe in detail how to evaluate statistical significance of
dependency patterns or their improvement under different sets of assumptions.

\subsection{Notations}
\label{subsec:prelnotations}

The mathematical notations used in this paper are given in Table
\ref{tab:notations}.  We note that the sample space spun by variables
$A_1,\dots,A_k$ is $\sspace=\Dom(A_1) \times\dots \times
\Dom(A_k)$. When $A_i$s are binary variables
$\sspace=\{0,1\}^k$. Sample points $\row\in \sspace$ correspond to
atomic events and all other events can be presented as their
disjunctions. Often these can be presented in a reduced form, for
example, event $(A_1{=}1,A_2{=}1)\vee(A_1{=}1,A_2{=}0)$ reduces to
$(A_1{=}1)$. In this paper, we focus on events that can be presented
as conjunctions $\Xisx$, where $\Xset \subseteq
\{A_1,\dots,A_k\}$. When it is clear from the context, we notate
elements of $\Xset$, $|\Xset|=m$, by $A_1,\hdots,A_m$ instead of more
complicated $A_{i_1},\hdots,A_{i_m}$, $\{i_1,\dots,i_m\}\subseteq
\{1,\dots,k\}$. We also note that data set $\data$ is defined as a
vector of data points so that duplicate rows (i.e., rows $\row_i$,
$\row_j$ where $\row_i=\row_j$ but $i\neq j$) can be distinguished by
their index numbers.

\begin{table}[!h]
\caption{Notations.}
\label{tab:notations}
\begin{center}
\begin{tabular}{ll}

${\mathbb{N}}={\mathbb{Z^+}}\cup \{0\}$&natural numbers (including 0)\\

$N_X$, $N_A$, $N_{XA}$, $N$ &random variables in $\mathbb{N}$\\

$A,B,C$,\ldots &variables\\

$\Dom(A)$&domain of variable $A$\\

$a_1,a_2,a_3,\ldots \in\Dom(A)$ &values of variable $A$\\

$A_1{=}a_1,A_2{=}a_2$&value assignment, where $A_1{=}a_1$ and $A_2{=}a_2$\\

$\Xset,\Yset,\Zset,\Qset$&variable sets (vector-valued variables)\\

$|\Xset|=m$ &number of variables in set $\Xset$ (its cardinality)\\

$\xbf=(a_1,\dots,a_m)$&vector of variable values\\

$(\Xisx)=((A_1{=}a_1),\dots,(A_l{=}a_m))$&value assignment of set $\Xset{=}\{A_1,...,A_m\}$, also interpreted as\\ 
&a conjunction of assignments $A_i{=}a_i$\\

$I_{\Xisx}$&an indicator variable for $\Xisx$; $I_\Xisx{=}1$, if $\Xisx$, and $I_\Xisx{=}0$\\ 
&otherwise\\

$A$, $\neg A$&short hand notations for $A{=}1$ and $A{=}0$, when $A$ is binary\\

$\Xset=((A{=}1),\hdots,(A_m{=}1))$&short hand notation for a conjunction of positive-valued assignments\\
$\neg\Xset=((A_1{=}0)\vee \dots \vee (A_m{=}0))$&and its complement when $\Xset$ is a set of binary variables\\

$\sspace=\Dom(A_i)\times \dots \times \Dom(A_k)$&sample space spun by variables $A_1,\hdots, A_k$\\

$\row=((A_1{=}a_1),\dots,(A_k{=}a_k)) \in \sspace$&a data point; corresponds to an atomic event in $\sspace$\\ 

$\data=(\row_1,\ldots,\row_n)$, $\row_i\in\sspace$&data set, a vector of data points, whose elements are also called\\ 
&rows or tuples of data\\

$|\data|=n$&number of data points in $\data$\\

$P(\Xisx)$&probability of event $\Xisx$\\

$\fr(\Xisx)$&absolute frequency of event $\Xisx$; number of data points where $\Xisx$\\

$\phi(\Xisx\rightarrow A{=}a)=P(A{=}a|\Xisx)$&precision of rule $\Xisx \rightarrow A{=}a$\\

$\gamma(\Xisx,A{=}a)=\frac{P(\Xisx,A{=}a)}{P(\Xisx)P(A{=}a)}$&lift of rule $\Xisx \rightarrow A{=}a$\\

$\delta(\Xisx,A{=}a)$&leverage of rule $\Xisx \rightarrow A{=}a$\\
$=P(\Xisx,A{=}a)-P(\Xisx)P(A{=}a)$&\\

$\Delta$, $\Gamma$& random variables for the leverage and lift\\

$H_0$, $H_A$&null hypothesis and an alternative hypothesis\\

$\{H_1,\dots, H_m\}$&a set of null hypotheses\\

$p$&probability value or a $p$-value\\

$p_i$&observed $p$-value of $H_i$\\

$P_i$&random variable for the $p$-value of hypothesis $H_i$\\

$p_F$&$p$-value defined by Fisher's exact test\\

$p_A$, $p_X$, $p_{XA}$, $p_{A|X}$&parameter values of discrete probability distributions\\

$\mathcal{M}$&statistical model, a `sampling scheme'\\

$\measure$&test statistic\\

$\measureval_i$&observed value of the test statistic of $H_i$\\

$\measure_i$&random variable for the value of the test statistic of $H_i$\\

$L(\cdot)$&likelihood function\\

$\MI(\cdot)$&mutual information\\

$z$&$z$-score\\

\end{tabular}
\end{center}
\end{table}

\subsection{Statistical dependence}
\label{subsec:stdep}

The notion of statistical dependence is equivocal and even the
simplest case, dependence between two events, is subject to
alternative interpretations.  Interpretations of statistical
dependence between more than two events or variables are even more
various. In the following, we introduce the main types of statistical
independence that are needed for defining dependency patterns and
evaluating their statistical significance and mutual relationships.

\subsubsection{Dependence between two events}

Definitions of statistical dependence are usually based on the
classical notion of statistical independence between two events. 
We begin from a simple case where the events are 
variable-value combinations, $\Aisa$ and $\Bisb$. 

\begin{definition}[Statistical independence between two events]
\label{statind2events} 
Let $\Aisa$ and $\Bisb$ be two events, $P(\Aisa)$ and $P(\Bisb)$ their 
marginal probabilities, and $P(\Aisa,\Bisb)$ their joint probability. 
Events $(\Aisa)$ and $(\Bisb)$ are {\em statistically independent}, if 
\begin{equation}
\label{eqindependence}
P(\Aisa,\Bisb)=P(\Aisa)P(\Bisb).
\end{equation}
\end{definition}

Statistical dependence is seldom defined formally, but in practice,
there are two approaches. If dependence is considered as a Boolean
property, then any departure from complete independence 
(Eq.~\eqref{eqindependence}) is defined as dependence. Another approach,
prevalent in statistical data analysis, is to consider dependence as a
continuous property ranging from complete independence to complete
dependence. Complete dependence itself is an ambiguous term, but
usually it refers to equivalence of events:
$P(\Aisa,\Bisb)=P(\Aisa)=P(\Bisb)$ (perfect positive dependence) or
mutual exclusion of events: $P(\Aisa,\Bneqb)=P(\Aisa)=P(\Bneqb)$
(perfect negative dependence).

The strength of dependence between two events can be evaluated with several 
alternative measures. In pattern discovery, two of the most popular measures 
are {\em leverage} and {\em lift}. 

Leverage is  equivalent to Yule's $\delta$ \citep{yule1912}, Piatetsky-Shapiro's
unnamed measure \citep{piatetskyshapiro}, and Meo's `dependence
value' \citep{meo}). It measures the absolute
deviation of the joint probability from its expectation under
independence:
\begin{equation}
\delta(\Aisa,\Bisb)=P(\Aisa,\Bisb)-P(\Aisa)P(\Bisb).
\end{equation}
We note that this is the same as covariance between binary variables
$A$ and $B$. 

Lift has also been called `interest'
\citep{brinmotwani}, `dependence' \citep{wuzhangzhang}, and `degree
of independence' \citep{yaozhong}). It measures the ratio of the joint probability 
and its expectation under
independence:
\begin{equation}
\gamma(\Aisa,\Bisb)=\frac{P(\Aisa,\Bisb)}{P(\Aisa)P(\Bisb)}.
\end{equation}

For perfectly independent events, leverage is $\delta=0$ and lift is
$\gamma=1$, for positive dependencies $\delta>0$ and $\gamma>1$, and
for negative dependencies, $\delta<0$ and $\gamma<1$. 

If the real probabilities of events were known, the strength of
dependence could be determined accurately. However, in practice, the
probabilities are estimated from the data. The most common method is
to approximate the real probabilities with relative frequencies
(maximum likelihood estimates) but other estimation methods are also
possible. The accuracy of these estimates depends on how
representative and error-free the data is. The size of the data
affects also precision, because continuous probabilities are
approximated with discrete frequencies. Therefore, it is quite
possible that two independent events express some degree of dependence
in the data (i.e., $\Pest(\Aisa,\Bisb)\neq\Pest(\Aisa)\Pest(\Bisb)$,
where $\Pest$ is the estimated probability, even if
$P(\Aisa,\Bisb)=P(\Aisa)P(\Bisb)$ in the population). In the worst
case, two events always co-occur in the data, indicating maximal
dependence, even if they are actually independent.  To some extent the
probability of such false discoveries can be controlled by statistical
significance testing, which is discussed in Subsection
\ref{subsec:sigtest}. In the other extreme, two dependent events may
appear independent in the data (i.e.,
$\Pest(\Aisa,\Bisb)=\Pest(\Aisa)\Pest(\Bisb)$). However, this is not
possible if the actual dependence is sufficiently strong (i.e.,
$P(\Aisa,\Bisb)=P(\Aisa)$ or $P(\Aisa,\Bisb)=P(\Bisb)$), assuming that
the data is error-free. Such missed discoveries are harder to detect,
but to some extent the problem can be alleviated by using {\em
  powerful} methods in significance testing (Subsection
\ref{subsec:sigtest}).

\subsubsection{Dependence between two variables}

For each variable, we can define several events which describe its
values. If the variable is categorical, it is natural to consider each
variable-value combination as a possible event. Then, the independence
between two categorical variables can be defined as follows:

\sloppy\begin{definition}[Statistical independence between two variables]
Let $A$ and $B$ be two categorical variables, whose domains 
are $\Dom(A)$ and $\Dom(B)$. $A$ and $B$ are {\em statistically
  independent}, if for all $a \in Dom(A)$ and $b\in Dom(B)$ 
$P(A{=}a,B{=}b)=P(A{=}a)P(B{=}b)$.
\end{definition}

Once again, dependence can be defined either as a Boolean property (lack of 
independence) or a continuous property.  However, there is no standard way to 
measure the strength of dependence between variables. In practice, the measure is
selected according to data and modelling purposes. Two commonly used
measures are the {\em $\chi^2$-measure}

\begin{equation}
\label{eqchi2general}
\chi^2(A,B)=\sum_{a\in \Dom(A)}\sum_{b\in \Dom(B)}\frac{n(P(A{=}a,B{=}b)-P(A{=}a)P(B{=}b))^2}{P(A{=}a)P(B{=}b)}
\end{equation}
and {\em mutual information}
\begin{equation}
\label{eqMIgeneral}
\MI(A,B)=\sum_{a\in \Dom(A)}\sum_{b\in \Dom(B)}P(A{=}a,B{=}b)log\frac{P(A{=}a,B{=}b)}{P(A{=}a)P(B{=}b)}.
\end{equation}

\begin{figure}[!h]
\begin{center}
\includegraphics[width=0.55\textwidth]{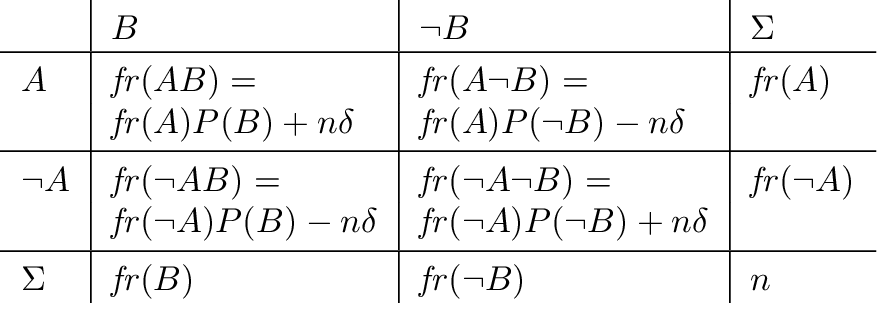}
\caption{A contingency table for two binary variables $A$ and $B$ expressing 
absolute frequencies of events $AB$, $A\neg B$, $\neg AB$ and $\neg A\neg B$ 
using leverage, $\delta=\delta(A,B)$.}
\label{conttablebin}
\end{center}
\end{figure}

\begin{figure}[!h]
\begin{center}
\includegraphics[width=0.9\textwidth]{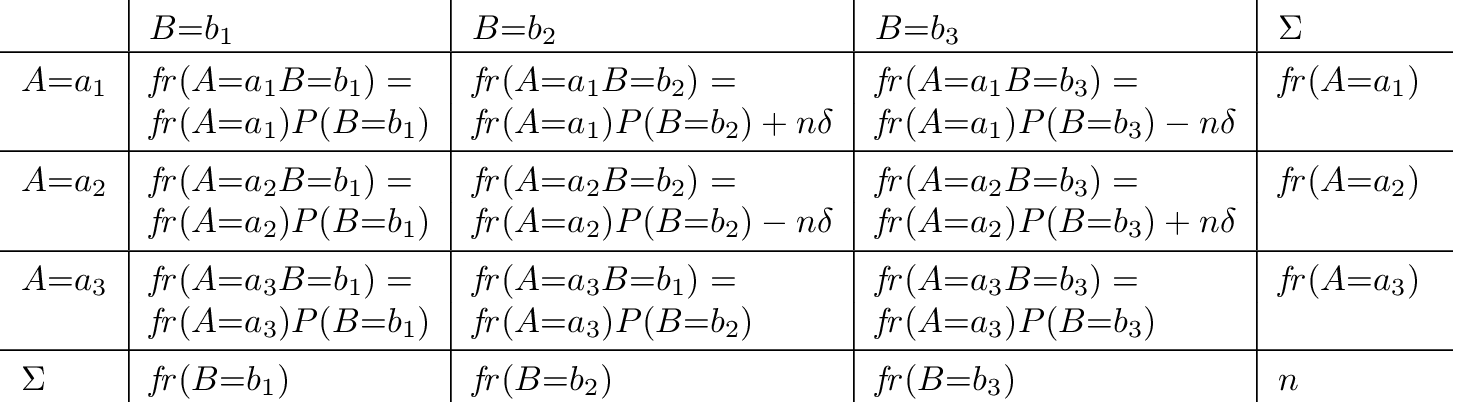}
\caption{An example contingency table where some value combinations of $A$ and $B$ express independence and others dependence. The frequencies are expressed using leverage, $\delta=\delta(A{=}a_1,B{=}b_2)$.}
\label{conttable3vl}
\end{center}
\end{figure}

If the variables are binary, the notions of independence between
variables and the corresponding events coincide. Now independence 
between any of the four value
combinations $AB$, $A\neg B$,$\neg AB$, $\neg A\neg B$ means
independence between variables $A$ and $B$ and vice versa. In
addition, the absolute value of leverage is the same for all value
combinations and can be used to measure the strength of dependence
between binary variables. This is shown in the corresponding
contingency table (Fig.~\ref{conttablebin}). Unfortunately, this
handy property does not hold for multivalued variables. Figure
\ref{conttable3vl} shows an example contingency table for two
three-valued variables where some value combinations are independent
and others dependent.

\subsubsection{Dependence between many events or variables}

The notion of statistical independence can be generalized to three or
more events or variables in several ways. The most common types of
independence are mutual independence, bipartition independence,
and conditional independence (see e.g.,
\citep[p.~318]{agrestibook}). In the following, we give general
definitions for these three types of independence.

In statistics and probability theory, mutual independence of a set of events is classically defined as follows (see e.g., \citep[p.~128]{feller}):

\begin{definition}[Mutual independence]
\label{defmutindep}
Let $\Xset=\{A_1,\hdots,A_m\}$ be a set of variables, whose domains are 
$\Dom(A_i)$, $i=1,\hdots,m$. Let $a_i\in \Dom(A_i)$ notate a value of $A_i$. 
A set of events $(A_1{=}a_1,\hdots,A_m{=}a_m)$ is called mutually independent if for 
all $\{i_1,\hdots,i_{m'}\}\subseteq \{1,\hdots,m\}$ holds 

\begin{equation}
P(A_{i_1}{=}a_{i_1},\hdots,A_{i_{m'}}{=}a_{i_{m'}})=\prod_{j=1,\hdots,m'}P(A_{i_j}{=}a_{i_j}).
\end{equation}
\end{definition}

If variables $A_i\in \Xset$ are binary, the conjunction of true-valued variables 
$(A_1{=}1,\hdots,$ $A_m{=}1)$ can be expressed as $A_1,\hdots,A_m$ and the
condition for mutual independence reduces to $\forall \Yset\subseteq \Xset$ 
$P(\Yset)=\prod_{A_i\in \Yset}P(A_i)$. An equivalent condition is to require
that for all truth value combinations $(a_1,\hdots,a_m)\in \{0,1\}^m$
holds $P(A_1{=}a_1,\hdots,A_m{=}a_m)=\prod_{i=1}^m P(A_i{=}a_i)$
\citep[p.~128]{feller}. We note that in data mining this property has sometimes been called independence of binary variables and independence of events has referred to a weaker condition (e.g., \citep{silverstein})

\begin{equation}
\label{weakercondition}
P(\Xset)=\prod_{A_i\in \Xset}P(A_i). 
\end{equation}
The difference is that in the latter it is not required that all
$\Yset\subsetneq \Xset$ should express independence. Both definitions
have been used as a starting point to define interesting set-formed
dependency patterns (e.g., \citep{Webb10,silverstein}). In this paper 
we will call this type of patterns {\em dependency sets} (Section
\ref{sec:sets}).

In addition to mutual independence, a set of events or variables can
express independence between different partitions of the
set. The only difference to the basic definition of statistical
independence is that now single events or variables have been replaced
by sets of events or variables. In this paper we call this type of independence bipartition independence.

\begin{definition}[Bipartition independence]
\label{defbipindep}
Let $\Xset$ be a set of variables. For any partition $\Xset{=}\Yset\cup \Zset$, 
where $\Yset\cap \Zset{=}\emptyset$, possible value combinations are 
notated by $\ybf\in \Dom(\Yset)$ and $\zbf\in\Dom(\Zset)$. 
\begin{itemize}
\item[(i)] Event $\Yisy$ is independent of event $\Zisz$, if 
$P(\Yisy,\Zisz)=P(\Yisy)P(\Zisz)$.
\item[(ii)] Set of variables $\Yset$ is independent of $\Zset$, if 
$P(\Yisy,\Zisz)=P(\Yisy)P(\Zisz)$ for all $\ybf\in\Dom(\Yset)$ and 
$\zbf\in\Dom(\Zset)$.
\end{itemize}
\end{definition}

Now one can derive a large number of different 
dependence patterns from a single set $\Xset$ or event $\Xisx$. There
are $2^{m-1}-1$ ways to partition set $\Xset$, $|\Xset|=m$, into two
subsets $\Yset$ and $\Zset=\Xset\setminus \Yset$ ($|\Yset|=1,\hdots,\lceil\frac{m-1}{2}\rceil$). In data mining, patterns expressing bipartition dependence 
between sets of events are often expressed as {\em dependency rules}
$\Yisy\rightarrow \Zisz$. Because both the rule antecedent and
consequent are binary conditions, the rule can be interpreted as  
dependence between two new binary (indicator) variables $I_\Yisy$ and $I_\Zisz$ 
($I_\Yisy{=}1$ if $\Yisy$ and $I_\Yisy{=}0$ otherwise). In statistical terms, this is the 
same as collapsing a multidimensional contingency
table into a simple $2\times 2$ table. In addition to statistical
dependence, dependency rules are often required to fulfil other
criteria like sufficient frequency, strength of dependency or
statistical significance. Corresponding patterns
between sets of variables are less often studied, because the search is
computationally much more demanding. In addition, collapsed
contingency tables can reveal interesting and statistically
significant dependencies between composed events, when no significant
dependencies could be found between variables.

The third main type of independence is conditional independence
between events or variables:

\begin{definition}[Conditional independence]
Let $\Xset$ be a set of variables. For any partition 
$\Xset=\Yset\cup \Zset\cup \Qset$, where 
$\Yset\cap \Zset=\emptyset$, $\Yset\cap \Qset=\emptyset$, 
$\Zset\cap \Qset=\emptyset$, possible value combinations are notated by
$\ybf\in \Dom(\Yset)$, $\zbf\in \Dom(\Zset)$, and
$\qbf\in\Dom(\Qset)$.
\begin{itemize}
\item[(i)] Events $\Yisy$ and $\Zisz$ are conditionally independent given 
$\Qisq$, if\\ 
$P(\Yisy,\Zisz\mid\Qisq)=P(\Yisy\mid\Qisq)P(\Zisz\mid\Qisq)$.
\item[(ii)] Sets of variables $\Yset$ and $\Zset$ are conditionally independent given 
$\Qset$, if\\ 
$P(\Yisy,\Zisz\mid\Qisq)=P(\Yisy\mid\Qisq)P(\Zisz\mid\Qisq)$\\
for all $\ybf\in \Dom(\Yset)$, $\zbf\in \Dom(\Zset)$, and $\qbf\in \Dom(\Qset)$.
\end{itemize}
\end{definition}

Conditional independence can be defined also for more than two sets of
events or variables, given a third one. For example, in set
$\{A,B,C,D\}$ we can find four conditional independencies given $D$:
$A \perp BC$, $B\perp AC$, $C\perp AB$, and $A\perp B \perp C$. 
However, these types of independence are seldom needed in practice. In 
pattern discovery notions of conditional independence and dependence between
events are used for inspecting improvement of a dependency rule
$\Yset\Qset\rightarrow C$ over its generalization $\Yset\rightarrow C$ (Section 
\ref{sec:redsig}). 
In machine learning conditional independence between variables or sets 
of variables is an important property for constructing full probability 
models, like Bayesian networks or log-linear models.

\subsection{Statistical significance testing}
\label{subsec:sigtest}

Often when searching dependency rules and sets the aim is to find
dependencies that hold in the population from which the sample is
drawn (cf.\ Fig.~\ref{fig:platon}). Statistical significance tests are
the tools that have been created to control the risk that such
inferences drawn from sample data do not hold in the population. This
subsection introduces the key concepts that underlie significance testing
and gives an overview of the main approaches that can be applied in
testing dependency rules and sets. The same principles can be applied
in testing other types of patterns, but a reader would be well advised
to consult a statistician with regard to which tests to apply and how
to apply them.

\begin{figure}[!h]
\begin{center}
\includegraphics[width=0.8\textwidth]{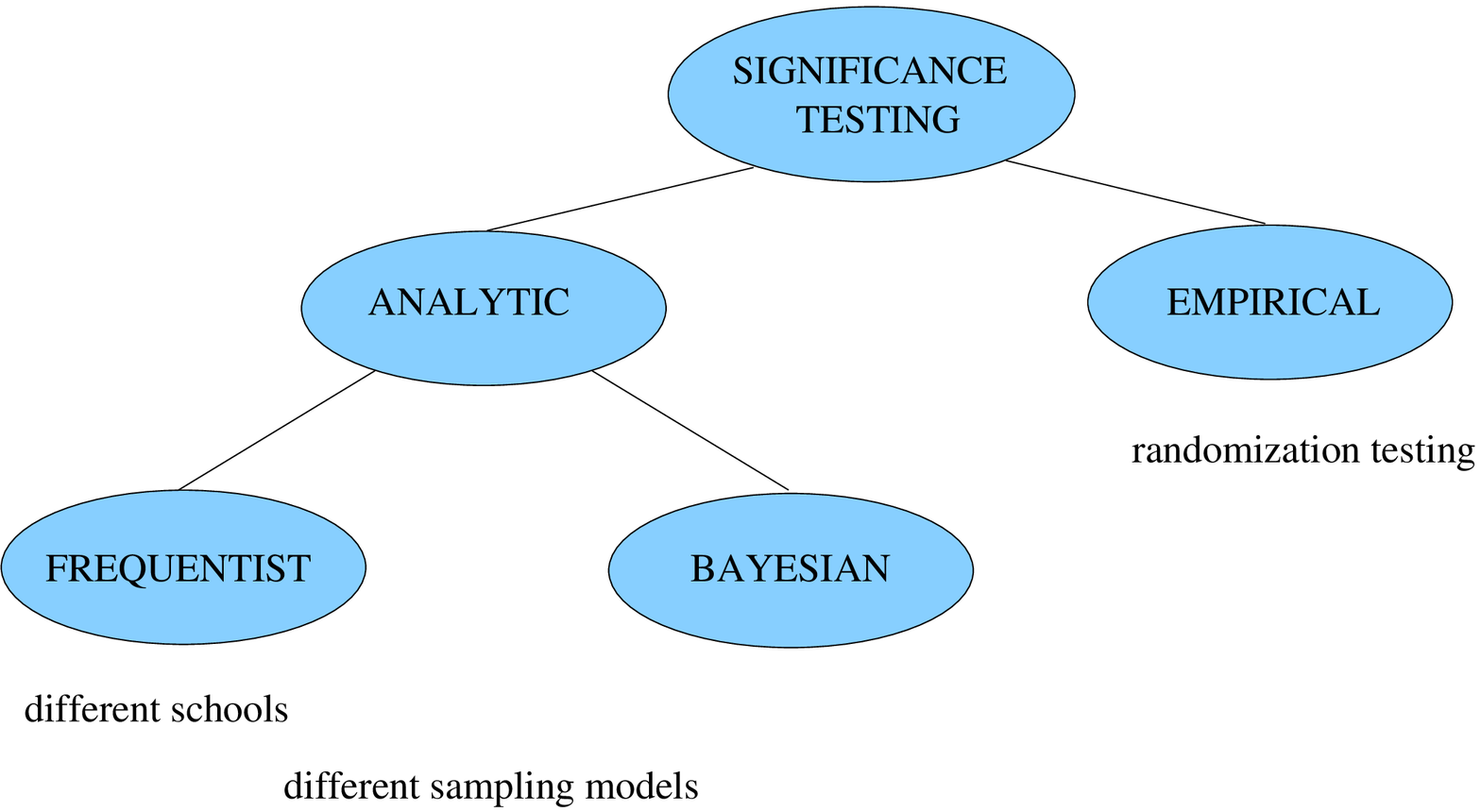}
\caption{Different approaches to statistical significance testing.}
\label{fig:approaches}
\end{center}
\end{figure}

The main idea of statistical significance testing is to estimate the
probability that the observed discovery would have occurred by chance. 
If the probability is very small, we can assume that the discovery is
genuine. Otherwise, it is considered spurious and discarded. The
probability can be estimated either analytically or empirically. The
analytical approach is used in the {\em traditional significance
  testing}, while {\em randomization tests} estimate the probability
empirically. Traditional significance testing can be further divided
into two main classes: the {\em frequentist} and {\em Bayesian
  approaches}. These main approaches to statistical significance
testing are shown in Fig.~\ref{fig:approaches}.

\subsubsection{Frequentist approach}

The frequentist approach of significance testing is the most commonly
used and best studied (see e.g.\ \citep[Ch.~26]{freedman} or
\citep[Ch.~10.1]{lindgren}). The approach is actually divided into two
opposing schools, Fisherian and Neyman-Pearsonian, but most textbooks
present a kind of synthesis (see e.g.,
\citep{hubbardbayarri2003}). The main idea is to estimate the
probability of the observed or a more extreme phenomenon $O$ under
some {\em null hypothesis}, $H_0$.  In general, the null hypothesis is
a statement on the value of some statistic or statistics $S$ in the
population. For example, when the objective is to test the
significance of dependency rule $\Xset \rightarrow A$, the null
hypothesis $H_0$ is the independence assumption:
$N_{XA}=nP(\Xset)P(A)$, where $N_{XA}$ is a random variable for the
absolute frequency of $\Xset A$. (Equivalently, $H_0$ could be
$\Delta=0$ or $\Gamma=1$, where $\Delta$ and $\Gamma$ are random
variables for the leverage and lift.) In independence testing the
null hypothesis is usually an equivalence statement, $S{=}s_0$
(nondirectional hypothesis), but in other contexts it can also be of
the form $S\leq s_0$ or $S\geq s_0$ (directional hypothesis). Often,
one also defines an explicit alternative hypothesis, $H_A$, which can
be either directional or nondirectional. For example, in pattern
discovery dependency rules $\Xset \rightarrow A$ are assumed to
express positive dependence, and therefore it is natural to form a
directional hypothesis $H_A$: $N_{XA}>nP(\Xset)P(A)$ (or $\Delta>0$ or
$\Gamma>1$).

\begin{figure}[!h]
\begin{center}
\includegraphics[width=0.6\textwidth]{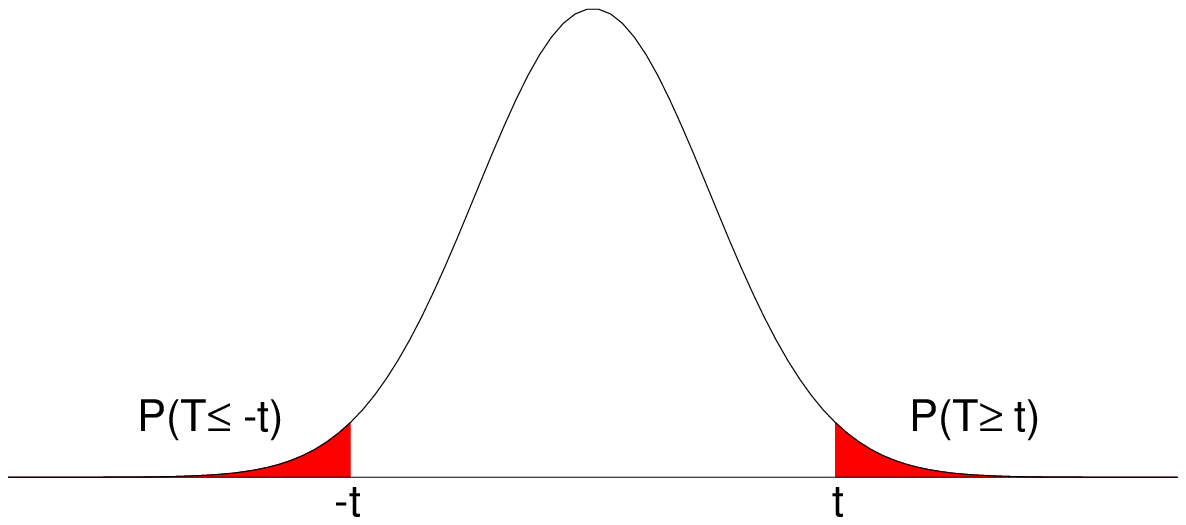}
\caption{An example distribution of test statistic $T$ under the null hypothesis. If the observed value of $T$ is $t$, the $p$-value is probability $P(T\geq t)$ (directional hypothesis) or $P(T\leq -t \vee T\geq t)$ (non-directional hypothesis).}
\label{fig:Tandp}
\end{center}
\end{figure}

When the null hypothesis has been defined, one should select a test statistic $\measure$ (possibly $S$ itself) and
define its distribution ({\em null distribution}) under $H_0$. The
$p$-value is defined from this distribution as the probability of the
observed or a more extreme $\measure$-value, $P(\measure\geq \measureval\mid H_0)$, $P(\measure\leq \measureval\mid H_0)$, or $P(\measure\leq -\measureval \textrm{ or }\measure\geq \measureval\mid H_0)$ (Fig.~\ref{fig:Tandp}). In the case
of independence testing, possible test statistics are, for example,
leverage, lift, and the $\chi^2$-measure (Eq.~\eqref{eqchi2general}). The distribution under independence is defined
according to the selected {\em sampling model}, which we will
introduce in Section \ref{sec:deprules}. The probability of observing
positive dependence whose strength is at least $\delta(\Xset,A)$ is
$P_{\Model}(\Delta\geq \delta(\Xset,A)\mid H_0)$, where $P_{\Model}$ is the
complementary cumulative distribution function for the assumed sampling model
$\Model$.

Up to this point, all frequentist approaches are more or less in
agreement. The differences appear only when the $p$-values are
interpreted. In the classical (Neyman-Pearsonian) hypothesis testing,
the $p$-value is compared to some predefined threshold $\alpha$. If
$p\leq \alpha$, the null hypothesis is rejected and the discovery is
called significant at level $\alpha$. Parameter $\alpha$ (also known
as the {\em test size}) defines the probability of committing a {\em type I
  error}, i.e., accepting a spurious pattern (and rejecting a correct
null hypothesis). Another parameter, $\beta$, is used to define the
probability of committing a {\em type II error}, i.e., rejecting a
genuine pattern as non-significant (and keeping a false null
hypothesis). The complement $1-\beta$ defines the {\em power} of the
test, i.e., the probability that a genuine pattern passes the
test. Ideally, one would like to minimize the test size and maximize
its power. Unfortunately, this is not possible, because $\beta$
increases when $\alpha$ decreases and vice versa. As a solution it
has been recommended (e.g.,\citep[p.~57]{lehmann}) to select appropriate
$\alpha$ and then to check that the power is acceptable given the
sample size. However, the power analysis can be difficult and all too
often it is skipped altogether.

The most controversial problem in hypothesis testing is how to select
an appropriate significance level. A convention is to use always the
same standard levels, like $\alpha{=}0.05$ or $\alpha{=}0.01$. However,
these values are quite arbitrary and widely criticized
(see e.g., (\citealp[p.~57]{lehmann}; \citealp{lecoutre2001}; \citealp{johnson1999})).
Especially in large data sets, the $p$-values tend to be very small and hypotheses get too easily rejected with conventional thresholds.
A simple alternative is to report only $p$-values, as advocated by
Fisher and also many recent statisticians (e.g., (\citealp[pp. 63-65]{lehmann}; \citealp{hubbardbayarri2003})). Sometimes,
this is called `significance testing' in distinction from `hypothesis
testing' (with fixed $\alpha$s), but the terms are not used
systematically. Reporting only $p$-values may often be sufficient, but
there are still situations where one should make concrete decisions
and a binary judgement is needed.

Deciding threshold $\alpha$ is even harder in data mining where
numerous patterns are tested. For example, if we use threshold
$\alpha{=}0.05$, then there is up to 5\% chance that a spurious
pattern passes the significance test. If we test 10 000 spurious
patterns, we can expect up to 500 of them to pass the test
erroneously.  This so called {\em multiple testing problem} is
inherent in knowledge discovery, where one often performs an
exhaustive search over all possible patterns. We will return to this
problem in Section \ref{sec:multtest}.

\subsubsection{Bayesian approach}

Bayesian approaches are becoming increasingly popular in both
statistics and data mining (see e.g., \citep{baeysiantutorial2016}).
However, to date there has been little uptake of them in statistically
sound pattern discovery. We include here a brief summary of the
Bayesian approach for completeness and in the hope that it will
stimulate further investigation of this promising approach.

The idea of Bayesian significance testing (see e.g., 
(\citealp[Ch.~4]{bayesstat}; \citealp{albert1997}; \citealp{jamil2017})) 
is quite similar to the frequentist approach, but
now we assign some prior probabilities $P(H_0)$ and $P(H_A)$ to the
null hypothesis $H_0$ and the alternative research hypothesis
$H_A$. Next, the conditional probabilities, $P(O\mid H_0)$ and $P(O\mid H_A)$,
of the observed or a more extreme phenomenon $O$ under $H_0$ and $H_A$
are estimated from the data. Finally, the probabilities of both
hypotheses are updated by the Bayes' rule and the acceptance or
rejection of $H_0$ is decided by comparing posterior probabilities
$P(H_0\mid O)$ and $P(H_A\mid O)$. The resulting conditional probabilities
$P(H_0\mid O)$ are asymptotically similar (under some assumptions even
identical) to the traditional $p$-values, but Bayesian
testing is sensitive to the selected prior probabilities
\citep{bayesiantesting}. One attractive feature of the Bayesian approach is that 
it allows to quantify the evidence for and against the null hypothesis. However, 
the procedure tends to be more complicated than the frequentist one; specifying prior distributions may require  
a plethora of parameters and the posterior probabilities cannot always be evaluated analytically \citep{agrestihitchcock,jamil2017}. 

\subsubsection{Randomization testing}

Randomization testing (see e.g., \citep{edgington}) offers a
relatively assumption-free approach for testing statistical
dependencies. Unlike traditional significance testing, there is no
need to assume that the data would be a random sample from the
population or to define what type of distribution the test statistic has
under the null hypothesis. Instead, the significance is estimated
empirically, by generating random data sets under the null hypothesis
and checking how often the observed or a more extreme phenomenon
occurs in them.

When independence between $A$ and $B$ is tested, the null hypothesis
is {\em exchangeability} of the $A$-values on rows when $B$-values are
kept fixed, or vice versa. This is the same as stating that all
permutations of $A$-values in the data are equally likely. A similar
null hypothesis can be formed for mutual independence in a set of variables.
 If only a single dependency set $\Xset$ is tested, it is enough to
generate random data sets $\data_1,\hdots,\data_b$ by permuting values
of each $A_i$, $A_i\in \Xset$.
Usually, it is required that all marginal probabilities $P(A_i)$
remain the same as in the original data, but there may be additional
constraints, defined by the {\em permutation scheme}. Test statistic
$\measure$ that evaluates goodness of the pattern is calculated in
each random data set.  For simplicity, we assume that the test
statistic $\measure$ is increasing by goodness (a higher value
indicates a better pattern). If the original data set produced
$\measure$-value $\measureval_0$ and $b$ random data sets produced
$\measure$-values $\measureval_1,\hdots,\measureval_b$, the empirical
$p$-value of the observed pattern is
\begin{equation}
p_{em}{=}\frac{|\{\data_j\mid\measureval_j\geq \measureval_0, j=1,\hdots,b\}|+1}{b+1}.
\end{equation}

If the data set is relatively small and $\Xset$ is simple, it is
possible to enumerate all possible permutations where the marginal
probabilities hold. This leads to an {\em exact permutation test},
which gives an exact $p$-value. On the other hand, if the data set is
large and/or $\Xset$ is more complex, all possibilities cannot be
checked, and the empirical $p$-value is less accurate. In this case,
the test is called a {\em random permutation test} or an {\em
  approximate permutation test}. There are also some special cases,
like testing a single dependency rule, where it is possible to express
the permutation test in a closed form that is easy to evaluate exactly
(see Fisher's exact test in Subsection \ref{subsec:varbased}).

An advantage of randomization testing is
that the test statistic can have any kind of distribution, which is
especially handy when the statistic is new or poorly known.
With randomization one can test also such
null hypotheses for which no closed form test exists.
Randomization tests are technically valid even if the data are not a random sample because strictly speaking the population to which the null hypotheses relate is the set of all permutations of the sample defined by the permutation scheme. 
However, the results can be generalized to the
reference population only to the extent of how representative the sample
was for that population \citep[p. 24]{legendre}.  
One critical problem with randomization testing is that it is not always clear 
how the data should be permuted, and different permutation schemes can produce quite different results in their assessment of patterns (see e.g.,\ \citep{hanhijarvi2011}).
The number of random permutations plays also an important role in
testing. The more random permutations are performed, the more accurate
the empirical $p$-values are, but in practice, extensive permuting can
be too time consuming. Computational costs restrict also the use of
randomization testing in search algorithms especially in large data
sets.

The idea of randomization tests can be extended for estimating the
overall significance of all mining results or even for tackling the
multiple testing problem. For example, one may test the significance
of the number of all frequent sets (given a minimum frequency
threshold) or the number of all sufficiently strong pair-wise
correlations (given a minimum correlation threshold) using
randomization tests \citep{gionismannila}. In this case, it is
necessary to generate complete data sets randomly for testing. The
difficulty is to decide what properties of the original data set
should be maintained. One common solution in pattern mining is to keep
both the column margins ($\fr(A_i)$s) and the row margins (numbers of
1s on each row) fixed and generate new data sets by {\em swap
  randomization} \citep{swaprand}. A prerequisite for this method is
that the attributes are semantically similar (e.g.\ occurrence or
absence of species) and it is sensible to swap their values. In
addition, there are some pathological cases, where no or only a few
permutations exist with the given row and column margins, resulting in
a large $p$-value, even if the original data set contains a
significant pattern.\citep{gionismannila}

\section{Statistical significance of dependency rules}
\label{sec:deprules}

Dependency rules are a famous pattern type that expresses 
bipartition dependence between the rule antecedent and the consequent.
In this section, we discuss how statistical significance of dependency
rules is evaluated under different assumptions. Especially, we
contrast two alternative interpretations of dependency rules that are
called variable-based and value-based interpretations and introduce
appropriate tests in different sampling models.

\subsection{Dependency rules}
\label{subsec:dependencyrules}

Dependency rules are maybe the simplest type of statistical dependency
patterns. As a result, it has been possible to develop efficient exhaustive
search algorithms. With these, dependency rules can reveal arbitrarily
complex bipartite dependencies from categorical or discretized
numerical data without any additional assumptions. This makes
dependency rule analysis an attractive starting point for any data
mining task.
In medical science, for example, an important task is to search for
statistical dependencies between gene alleles, environmental factors,
and diseases. We recall that statistical dependencies are not necessarily causal
relationships, but still they can help to form causal hypotheses and
reveal which factors predispose or prevent diseases (see e.g., 
\citep{causalrules,LiCausal2015}). Interesting
dependencies do not necessarily have to be strong or frequent, but
instead, they should be statistically valid, i.e., genuine
dependencies that are likely to hold also in future data. 
In addition, it is often required 
that the patterns should not contain any superfluous variables which
would only obscure the real dependencies. Based on these
considerations, we will first give a general definition of dependency
rules and then discuss important aspects of genuine 
dependencies.

\begin{definition}[Dependency rule]
Let $\Rset$ be a set of categorical variables, $\Xset\subseteq \Rset$, and 
$\Yset \subseteq \Rset\setminus \Xset$. Let us denote value vectors of 
$\Xset$ and $\Yset$ by $\xbf \in \Dom(\Xset)$ and $\ybf \in \Dom(\Yset)$.
Rule $\Xisx \rightarrow \Yisy$ is a dependency rule, if 
$P(\Xisx,\Yisy)\neq P(\Xisx)P(\Yisy)$.

The dependency is
(i) positive, if $P(\Xisx,\Yisy)>
P(\Xisx)P(\Yisy)$, and
(ii) negative, if $P(\Xisx,\Yisy)<
P(\Xisx)P(\Yisy)$.
Otherwise, the rule expresses independence.
\end{definition}

It is important to recognize that while the convention is to specify the antecedent and consequent and use a directed arrow to distinguish them, statistical dependence is a symmetric relation and strictly speaking the direction is arbitrary. Often, the rule is
expressed with the antecedent and consequent selected so that the precision 
('confidence') of the rule ($\phi(\Xisx \rightarrow \Yisy)=P(\Yisy\mid\Xisx)$ or
$\phi(\Yisy \rightarrow \Xisx)=P(\Xisx\mid\Yisy)$) is maximal. An exception is supervised descriptive rule discovery (including class association rules \citep{li2001cmar}, subgroup discovery \citep{herrera2011overview}, emerging pattern mining \citep{dong1999efficient} and contrast set mining \citep{bay2001detecting}),  where the consequent is fixed \citep{NovakLavracWebb09}.

For simplicity, we will concentrate on a common special case of
dependency rules where 1) all variables are binary, 2) the consequent
$\Yisy$ consists of a single variable-value combination, $A{=}i$,
$i\in\{0,1\}$, and 3) the antecedent $\Xisx$ is a conjunction of
true-valued attributes, i.e., $\Xisx \equiv (A_1{=}1,\hdots,A_l{=}1)$,
where $\Xset=\{A_1,\hdots,A_l\}$. With these restrictions the
resulting rules can be expressed in a simpler form 
$\Xset\rightarrow A{=}i$, where $i\in\{0,1\}$, or $\Xset\rightarrow A$ 
and $\Xset \rightarrow \neg A$.  Allowing negated consequents means that it is
sufficient to represent only positive dependencies (a positive
dependency between $\Xset$ and $\neg A$ is the same as a negative dependency
between $\Xset$ and $A$).  We note that this restriction is purely
representational and the following theory is easily extended to general dependency rules as well. Furthermore, we recall that this simpler form of rules can
still represent all dependency rules after suitable data transformations 
(i.e., creating new binary variables for all values of the original 
variables).

Finally, we note that dependency rules deviate from 
traditional {\em association rules} \citep{agrawalass} in their requirement of 
statistical dependence. Traditional association rules do not necessarily express any statistical dependence but relations between frequently occurring attribute sets. However, there has been research on association rules
where the requirement of minimum frequency ('minimum support') has been replaced by
requirements of statistical dependence (see e.g.,
\citep{webboptimal,webbcritical,webbml,kingfkais,stataprkais,jiuyongli,morishitasese,nijssen,nijssenguns}). For clarity, we will use here the term `dependency rule' for all rule type
patterns expressing statistical dependencies, even if they had been
called association rules, classification rules, or other similar patterns in the original
publications.

Statistical dependence is a necessary requirement of a dependency
rule, but in addition, it is frequently useful to impose further
constraints like that of statistical significance and absence of
superfluous variables. The following example illustrates some of these properties 
of dependency rules.

\begin{example}
\label{heartdiseaseexample}
Let us consider an imaginary database consisting of 1000 patients (50\% female, 50\% male), 30\% of them with heart disease. The database contains information on patients and their life style like smoking status, drinking coffee, having stress, going for sports, and using natural products.
Table \ref{tab:athero} lists some candidate dependency rules related to heart 
disease together with their frequency, precision, leverage, and lift.

\begin{table}[!h]
\caption{An imaginary example of dependency rules related to heart disease. $\fr$=frequency, 
$\phi$=precision, $\delta$=leverage, $\gamma$=lift.}
\label{tab:athero}
\begin{center}
\begin{tabular}{rlrrrr}
&rule&$\fr$&$\phi$&$\delta$&$\gamma$
{\rule{0pt}{2.2ex}}\\
\hline
1&smoking $\rightarrow$ heart disease&120&0.400&0.030&1.333{\rule{0pt}{2.6ex}}\\
2&sports $\rightarrow$ $\neg$ heart disease&400&0.800&0.050&1.143\\
3&coffee $\rightarrow$ $\neg$ heart disease&240&0.700&0.000&1.000\\
4&stress $\rightarrow$ heart disease&150&0.300&0.000&1.000\\
5&pine bark extract $\rightarrow$ $\neg$ heart disease&1&1.000&$<$0.001&1.429\\
6&female $\rightarrow$ $\neg$ heart disease&352&0.704&0.002&1.006\\
7&female, stress $\rightarrow$ heart disease&100&0.385&0.022&1.282\\
8&stress, smoking $\rightarrow$ heart disease&100&0.500&0.040&1.667\\
9&smoking, coffee $\rightarrow$ heart disease&96&0.400&0.024&1.333\\
10&smoking, sports $\rightarrow$ heart disease&20&0.333&0.020&1.111\\
11&female, sports $\rightarrow$ $\neg$ heart disease&203&0.808&0.027&1.154{\rule[-1.2ex]{0pt}{0pt}}\\
\end{tabular}
\end{center}
\end{table}

The first two rules are examples of simple positive and negative
dependencies (predisposing and protecting factors for heart
disease). Rules 3 and 4 are included as examples of so called
independence rules that express statistical independence between the
antecedent and consequent. Normally, such rules would be pruned out by
dependency rule mining algorithms.

Rule 5 is an example of a spurious rule, which is statistically
insignificant and likely due to chance. The database contains only one
person who uses pine bark extract regularly and who does not have
heart disease. Note that the lift is still quite large, the maximal
possible for that consequent. Rule 6 is also statistically insignificant, but for a different reason. The rule is very common, but the difference in the prevalence of heart disease among female and male patients is so small (148 vs.~152) that it can be explained by chance.

Rule 7 demonstrates non-monotonicity of statistical dependence. The
combination of stress and female gender correlates positively with
heart disease, even though stress alone was independent of heart disease
and the female gender was negatively correlated with it.

The last four rules illustrate the problem of superfluous
variables. In rule 8, neither of the condition attributes is superfluous,
because the dependency is stronger and more significant than simpler
dependencies involving only stress or only smoking. However, rules
9--11 demonstrate three types of superfluous rules where extra
factors i) have no effect on the dependency, ii) weaken it, or iii)
apparently improve it but not significantly. Rule 9 is superfluous,
because coffee has no effect on the dependency between smoking and
heart disease (coffee consumption and heart disease are conditionally
independent given smoking). Rule 10 is superfluous, because going for sports
weakens the dependency between smoking and heart disease. 
This kind of modifying effect might be interesting in some contexts,
if it were statistically significant. However, dependency rule mining
algorithms do not usually perform such analysis.  Rule 11 is the most
difficult to judge, because the dependence is itself significant and
the rule has larger precision and lift than either of simpler
dependencies involving only the female gender or only sports. However,
the improvement with respect to rule 2 is so small ($\phi=0.808$
vs.\ $\phi=0.800$) that it is likely due to chance.
\end{example}

In the previous example we did not state which measure should be preferred for measuring the strength of dependence or how the statistical significance should be evaluated. The reason is that {\em the selection of these measures as well as evaluation of 
statistical significance and superfluousness depend on the
  interpretation of dependency rules}. In principle there are two
alternative interpretations for rule $\Xset\rightarrow A{=}i$, $i\in\{0,1\}$:
either it can represent a dependency between events $\Xset$ (or $I_\Xset{=}1$) and 
$A{=}i$ or between variables $I_\Xset$ and $A$, where $I_\Xset$ is an indicator 
variable for event $\Xset$. These two interpretations have
sometimes been called {\em value-based} and {\em variable-based}
semantics \citep{blanchard} of the rule. Unfortunately, researchers
have often forgotten to mention explicitly which interpretation they
follow. This has caused much confusion and, in the worst case, led to missed
or inappropriate discoveries. The following example demonstrates how
variable- and value-based interpretations can lead to different
results.

\begin{example}
\label{ex:apple}
Let us consider a database of 100 apples describing their colour (green
or red), size (big or small), and taste (sweet or bitter).  Let us
notate $A${=}sweet, $\neg A${=}bitter (not sweet), $\Yset${=}\{red\}, $\neg \Yset{=}\{green\}$ (not red), $\Xset{=}\{red,big\}$ and $\neg \Yset{=}\neg \{red,big\}$ (i.e., green or small).  

\begin{figure}[!h]
\begin{center}
\includegraphics[width=0.6\textwidth]{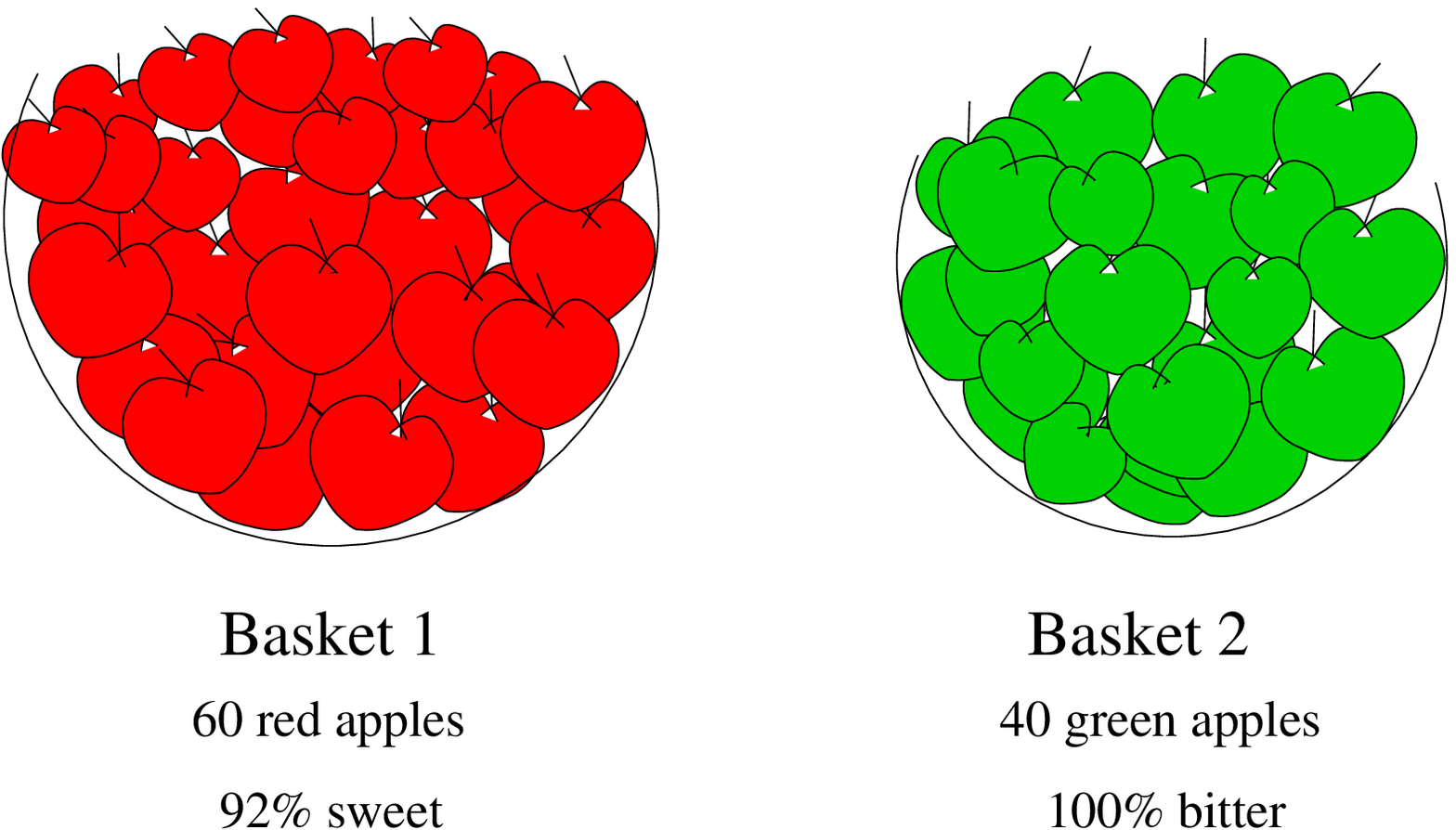}
\includegraphics[width=0.6\textwidth]{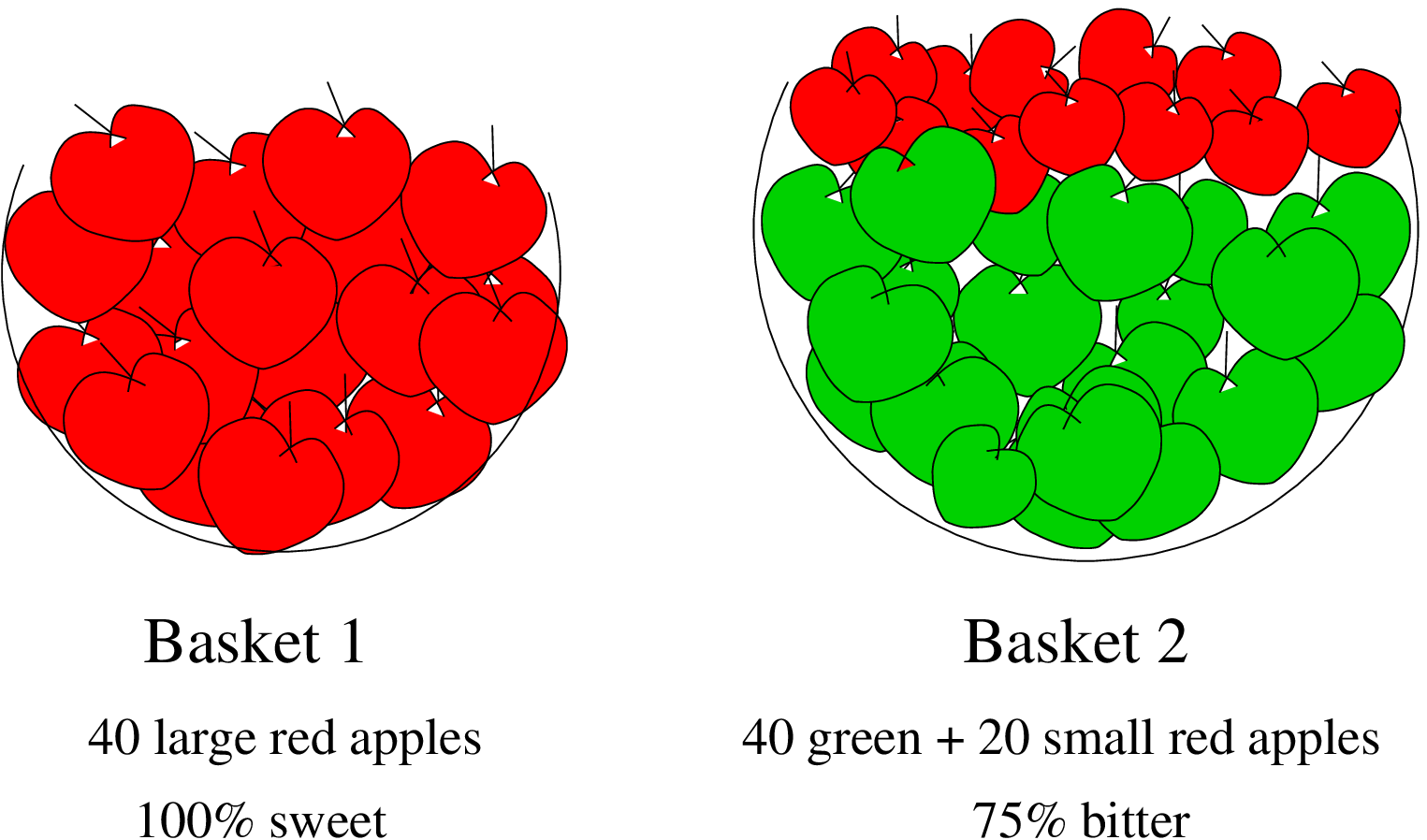}
\end{center}
\caption{Apple baskets corresponding to rules {\em red $\rightarrow$ sweet} (top) and {\em red and big $\rightarrow$ sweet} (bottom).}
\label{fig:apples}
\end{figure}

We would like to find strong dependencies related to either variable
'taste' (variable-based interpretation) or value 'sweet' (value-based
interpretation). Figure \ref{fig:apples} represents two such rules:
$\Yset\rightarrow A$ ({\em red $\rightarrow$ sweet}) and $\Xset\rightarrow A$
({\em red and big $\rightarrow$ sweet}).

The first rule expresses a strong dependency between binary variables 
$I_{\Yset}$ and $A$ (i.e., colour and taste) with $P(A|\Yset){=}0.92$, $P(\neg A|\neg \Yset){=}1.0$,
$\delta(\Yset,A){=}0.22$, and $\gamma(\Yset,A){=}1.67$.  So, with this rule we
can divide the apples into two baskets according to colour. The first
basket contains 60 red apples, 55 of which are sweet, and the second
basket contains 40 green apples, which are all bitter.  This is quite
a good rule if the goal is to classify well both sweet apples (for
eating) and bitter apples (for juice and cider).

The second rule expresses a strong dependency between the value
combination $\Xset$ (red and big) and value $A=1$ (sweet) with
$P(A|\Xset){=}1.0$, $P(\neg A|\neg \Xset){=}0.75$, $\delta(\Xset,A){=}0.18$,
$\gamma(\Xset,A){=}1.82$. This rule produces a basket of 40 big, red
apples, all of them sweet, and another basket of 60 green or small
apples, 45 of them bitter. This is an excellent rule if we would
like to predict sweetness better (e.g., get a basket of sweet apples
for our guests) without caring how well bitterness is predicted.

So, the choice between variable-based and value-based interpretation
results in a preference for a different rule. Either one can be
desirable for different modelling purposes.  This decision affects
also which goodness measure should be used. Leverage suits the
variable-based interpretation, because its absolute value is the same
for all truth value combinations ($\Xset A$, $\Xset\neg A$, $\neg
\Xset A$, $\neg \Xset\neg A$), but it may miss interesting
dependencies related to particular values. Lift, on the other hand,
suits the value-based interpretation, because it favours rules where
the given values are strongly dependent. However, it is not a reliable
measure alone, because it ranks well also coincidental `noise rules'
(e.g., {\em apple maggot $\rightarrow$ bitter}). Therefore, it has to
be accompanied with statistical significance tests.
\end{example}

In general, the variable-based interpretation tends to produce more
reliable patterns, in the sense that the discovered dependencies hold
well in future data (see e.g., \citep[Ch.~5]{whthesis}). However, there
are applications where the value-based interpretation may better
identify interesting dependency rules. 
One example could be analysis of predisposing factors (like gene alleles) 
for a serious disease. Some factors $\Xset$ may be rare, but still 
their occurrence could strongly predict the onset of  
some disease $D$. Medical scientists would
certainly want to find such dependencies $\Xset \rightarrow D$, even
if the overall dependency between variables $I_{\Xset}$ and $D$ would be
weak or insignificant.

In the following sections, we will examine how statistical
significance is tested in the variable-based and value-based
interpretations.

\subsection{Sampling models for the variable-based interpretation}
\label{subsec:varbased}

In the variable-based interpretation, the significance of dependency rule
$\Xset \rightarrow A$ is determined by classical independence
tests. The task is to estimate the probability of the observed or a
more `extreme' contingency table, assuming that variables $I_\Xset$ and $A$ were
actually independent. There is no consensus how the extremeness
relation should be defined, but intuitively, contingency table $\ctable_i$
is more extreme than table $\ctable_j$, if the dependence between $\Xset$
and $A$ is stronger in $\ctable_i$ than in $\ctable_j$. So, any measure for the
strength of dependence between variables can be used as a {\em
  discrepancy measure}, to order contingency tables. The simplest such
measure is leverage, but also {\em odds ratio}
\begin{equation}
odds(N_{X A},N_{X\neg A},N_{\neg X A},N_{\neg X\neg
  A})=\frac{N_{X A}N_{\neg X\neg A}}{N_{X\neg A}N_{\neg X A}}
\end{equation}
is commonly used. 
We note that odds ratio is not defined when $N_{X\neg A}N_{\neg X A}=0$ and some special policy is needed for these cases. 
In the following, we will notate the relation
``table $\ctable_i$ is equally or more extreme to table $\ctable_j$'' by $\ctable_i
\succeq \ctable_j$.

The probability of each contingency table $\ctable_i$ depends on the assumed
statistical model $\Model$. Model $\Model$ defines the space of all
possible contingency tables $\mathcal{T}_\Model$ (under the model
assumptions) and the probability $P(\ctable_i\mid\Model)$ of each table
$\ctable_i\in {\mathcal{T_M}}$. Because the task is to test independence, the
assumed model should satisfy the independence assumption
$P(\Xset A)=P(\Xset)P(A)$ in some form. For the probabilities
$P(\ctable_i\mid\Model)$ holds
$$\sum_{\ctable_i\in \mathcal{T}_\Model}P(\ctable_i\mid{\Model})=1.$$
 Now the probability of the observed contingency table $\obsctable$ or any 
$\ctable_i$, $\ctable_i\succeq \obsctable$, is the desired $p$-value

\begin{equation}
\label{pvalue}
p=\sum_{\ctable_i\succeq \obsctable}P(\ctable_i\mid{\Model}).
\end{equation}

Classically, statistical models for independence testing have been divided
into three main categories (sampling schemes)
\citep{barnard,pearson1947}, which we call {\em
  multinomial}, {\em double binomial}, and {\em hypergeometric}
models. In the statistics literature (e.g.\ \citep{barnard,upton}), the
corresponding sampling schemes are called {\em double dichotomy}, {\em
  2 $\times$ 2 comparative trial}, and {\em 2 $\times$ 2 independence
  trial}.

In the following we describe the three models using the classical urn
metaphor. However, because there are two binary variables of interest,
$I_\Xset$ and $A$, we cannot use the basic urn model with white and
black balls. Instead, we will use an apple basket model, with red and
green, sweet and bitter apples, like in Example \ref{ex:apple}.

\subsubsection*{Multinomial model}

In the multinomial model, it is assumed that the
real probabilities of sweet red apples, bitter red apples, sweet green
apples, and bitter green apples are defined by parameters $p_{X A}$,
$p_{X\neg A}$, $p_{\neg X A}$, and $p_{\neg X \neg A}$. The probability
of red apples is $p_X$ and of green apples $1-p_X$. Similarly, the
probability of sweet apples is $p_A$ and of bitter apples
$1-p_A$. According to the independence assumption, $p_{X A}=p_X p_A$,
$p_{X\neg A}=p_X(1-p_A)$, $p_{\neg X A}=(1-p_X)p_A$, and $p_{\neg X\neg
  A}=(1-p_X)(1-p_A)$. A sample of $n$ apples is taken {\em randomly}
from an infinite basket (or from a finite basket with replacement). Now the probability of obtaining $N_{X A}$ sweet red
apples, $N_{X\neg A}$ bitter red apples, $N_{\neg X A}$ sweet green
apples, and $N_{\neg X\neg A}$ bitter green apples is defined by
multinomial probability
\begin{multline}
\label{multinom}
P(N_{X A},N_{X\neg A},N_{\neg X A},N_{\neg X\neg A}\mid n,p_X,p_A)
=\left( n \atop N_{X A}, N_{X\neg A},N_{\neg X A}, N_{\neg X \neg
  A}\right)\\\cdot p_X^{N_X}(1-p_X)^{n-N_X}p_A^{N_A}(1-p_A)^{n-N_A}.
\end{multline}
\sloppy Since data size $n$ is given, the contingency tables can be defined by
triplets $\langle N_{X A},$ 
$N_{X\neg A},N_{\neg X A}\rangle$ or, equivalently, triplets
$\langle N_X, N_A, N_{X A}\rangle$. Therefore, the space of all possible contingency
tables is
\begin{multline*}
{\mathcal{T}_\Model}=\{\langle N_X,N_A,N_{X A}\rangle\mid N_X{=}0,\dots ,n; N_A{=}0,\dots ,n;
N_{X A}{=}0,\dots ,\min\{N_X,N_A\}\}.
\end{multline*}
For estimating the $p$-value with Eq.~\eqref{pvalue}, we should
still solve two problems. First, the parameters $p_X$ and $p_A$ are
unknown. The most common solution is to estimate them by the observed
relative frequencies (maximum likelihood estimates). 
Second, we should decide when a
contingency table $\ctable_i$ is equally or more extreme than the observed
contingency table $\obsctable$. For this purpose, we have to select the
discrepancy measure, which evaluates the overall dependence in a
contingency table, when only the data size $n$ is fixed. Examples of
such measures are leverage and the odds ratio.

In practice, the multinomial test is seldom used, but the multinomial
model is an important theoretical model, from which other models can
be derived as special cases.

\subsubsection*{Double binomial model}

In the double binomial model, it is 
assumed that we have two infinite baskets, one for red and one for green
apples. Let us call these the red and the green basket. In the red basket 
the probability of sweet apples is $p_{A|X}$ and of bitter apples
$1-p_{A|X}$, and in the green basket the probabilities are $p_{A|\neg X}$
and $1-p_{A|\neg X}$. According to the independence assumption, the
probability of sweet apples is the same in both baskets:
$p_A=p_{A|X}=p_{A|\neg X}$. A sample of $\fr(\Xset)$ apples is taken
randomly from the red basket and another random sample of $\fr(\neg \Xset)$
apples is taken from the green basket. The probability of obtaining
$N_{X A}$ sweet apples among the selected $\fr(\Xset)$ red apples is defined
by the binomial probability
\begin{equation*}
P(N_{X A}\mid\fr(\Xset),p_A)=\left(\fr(\Xset) \atop N_{X A}\right) p_A^{N_{X A}}(1-p_A)^{\fr(\Xset)-N_{X A}}.
\end{equation*}
Similarly, the probability of obtaining $N_{\neg X A}$ sweet apples
among the selected green apples is 
\begin{equation*}
P(N_{\neg X A}\mid\fr(\neg \Xset),p_A)=\left(\fr(\neg \Xset) \atop N_{\neg
  X A}\right) p_A^{N_{\neg X A}}(1-p_A)^{\fr(\neg \Xset)-N_{\neg X A}}.
\end{equation*}
Because the two samples are independent from each other, the 
probability of obtaining $N_{X A}$ sweet apples from $\fr(\Xset)$ red apples
and $N_{\neg X A}$ sweet apples from $\fr(\neg \Xset)$ green apples is the
product of the two binomials
\begin{equation}
\label{doublebin}
P(N_{X A},N_{\neg X A}\mid n,\fr(\Xset),p_A)=
\left(\fr(\Xset) \atop N_{X A}\right) \left(\fr(\neg \Xset) \atop N_{\neg
  X A}\right)p_A^{N_A}(1-p_A)^{n-N_A},
\end{equation}
where $N_A{=}N_{X A}+N_{\neg X A}$ is the total number of the obtained
sweet apples. (Here $\fr(\neg \Xset)$ was dropped from the condition, because $n$ is given.) We note that the double binomial probability is not
{\em exchangeable} with respect to the roles of $\Xset$
and $A$, i.e., generally
$P(N_{X A},N_{\neg X A}\mid n,\fr(\Xset),p_A)\neq P(N_{X A},N_{X\neg A}\mid n,\fr(A),p_X).$
In practice, this means that the probability of obtaining $\fr(\Xset A)$ sweet
red apples, $\fr(\Xset\neg A)$ bitter red apples, $\fr(\neg \Xset A)$ sweet green
apples, and $\fr(\neg \Xset\neg A)$ bitter green apples is (nearly always)
different in the model of the red and green baskets from the model of the
sweet and bitter baskets.

Since $\fr(\Xset)$ and $\fr(\neg \Xset)$ are given, each contingency table is
defined as a pair $\langle N_{X A},N_{\neg X A}\rangle$ or, equivalently,
$\langle N_{A},N_{X A}\rangle $. The space of all possible contingency tables is
$${\mathcal{T}_\Model}=\{\langle N_{X A},N_{\neg X A}\rangle \mid N_{X A}{=}0,\dots,\fr(\Xset);
N_{\neg X A}{=}0,\dots,\fr(\neg \Xset)\}.$$
We note that $N_A$ is not fixed, and therefore $N_A$ is generally not
equal to the observed $\fr(A)$. 

For estimating the significance with Equation \eqref{pvalue}, we should 
estimate the unknown parameter $p_A$ and select a discrepancy measure, like 
leverage or odds ratio. Then the exact $p$-value is obtained by summing over 
all possible values of $N_{XA}$ and $N_{\neg XA}$ 
where the dependence is sufficiently strong. However, often this is considered 
impractical and the $p$-value is approximated with asymptotic tests, which are 
discussed later.

\subsubsection*{Hypergeometric model}

In the hypergeometric model, there is no
sampling from an infinite basket. Instead, we can assume that we are given a
finite basket of $n$ apples, containing exactly $\fr(\Xset)$ red apples and
$\fr(\neg \Xset)$ green apples. We test all $n$ apples and find that $\fr(\Xset A)$
of red apples and $\fr(\neg \Xset A)$ of green apples are sweet. The
question is how probable is our basket, or the set of all at least
equally extreme baskets, among all possible apple baskets with $\fr(\Xset)$ red
apples, $\fr(\neg \Xset)$ green apples, $\fr(A)$ sweet apples, and $\fr(\neg A)$
bitter apples.

Now the baskets correspond to contingency tables. The number of all possible
baskets with the fixed totals $\fr(\Xset)$, $\fr(\neg \Xset)$, $\fr(A)$, and $\fr(\neg A)$ is 
\begin{equation*}
\sum_{i{=}0}^{\fr(A)} \left(\fr(\Xset) \atop i \right)\left(\fr(\neg \Xset) \atop \fr(A)-i
\right)=\left(n \atop \fr(A)\right).
\end{equation*}
(We recall that customarily $\left(m \atop l \right){=}0$, when
$l>m$.)
Assuming that all baskets with these fixed totals are equally likely, the
probability of a basket with $N_{X A}$ sweet red apples is 
\begin{equation*}
P(N_{X A}\mid n, \fr(\Xset),\fr(A))=\left(n \atop \fr(A)\right)^{-1}.
\end{equation*}

Because all totals are fixed, the extremeness relation is also easy to
define. Positive dependence is stronger than observed, when
$N_{X A}>\fr(\Xset A)$. For the $p$-value it is enough to sum the
probabilities of baskets containing at least $\fr(\Xset A)$ sweet red
apples. The resulting $p$-value is
\begin{equation}
\label{pfpos}
p_F=\sum_{i{=}0}^{J_1}\frac{\left(\fr(\Xset) \atop \fr(\Xset A)+i\right)\left(\fr(\neg
  \Xset) \atop \fr(\neg \Xset\neg A)+i\right)}{\left(n \atop \fr(A)\right)},
\end{equation}
where $J_1=\min\{\fr(\Xset\neg A), \fr(\neg \Xset A)\}$. (Instead of $J_1$ we could
give an upper range $\fr(A)$, because the zero terms disappear.) This
$p$-value is known as {\em Fisher's $p$}, because
it is used in {\em Fisher's exact test}, 
an exact permutation test. 
We give it a special symbol $p_F$, because it will be used later. For negative
dependence between red and sweet apples (or positive dependence
between green and sweet apples) the $p$-value is
\begin{equation}
\label{pfneg}
p_F=\sum_{i{=}0}^{J_2}\frac{\left(\fr(\Xset) \atop \fr(\Xset A)-i\right)\left(\fr(\neg
  \Xset) \atop \fr(\neg \Xset\neg A)-i\right)}{\left(n \atop \fr(A)\right)},
\end{equation}
where $J_2=\min\{\fr(\Xset A),\fr(\neg \Xset \neg A)\}$. 

\subsubsection*{Asymptotic measures}

We have seen that the $p$-values in the multinomial and double
binomial models are quite difficult to calculate. However, the
$p$-value can often be approximated easily using asymptotic
measures. With certain assumptions, the resulting $p$-values converge
to the correct $p$-values, when the data size $n$ (or $\fr(\Xset)$ and
$\fr(\neg \Xset)$) tend to infinity. In the following, we introduce two
commonly used asymptotic measures for independence testing: the
$\chi^2$-measure and mutual information. In statistics, the latter
corresponds to the {\em log likelihood ratio} \citep{neymanpearson1928}.

The main idea of asymptotic tests is that instead of estimating the
probability of the contingency table as such, we calculate some better
behaving test statistic $\measure$. If $\measure$ gets value $\measureval$, 
we estimate the probability of $P(\measure\geq \measureval)$ (assuming that 
large $\measure$-values indicate a strong dependency).

In the case of the $\chi^2$-test\index{$\chi^2$-test}, the test
statistic is the $\chi^2$-measure. Now the variables are binary and
Eq.~\eqref{eqchi2general} reduces into a simpler form:
\begin{align}
\label{chi2eq}
\chi^2=&\sum_{i{=}0}^1 \sum_{j{=}0}^1 
\frac{n(P(I_\Xset{=}i,A{=}j)-P(I_\Xset{=}i)P(A{=}j))^2}{P(I_\Xset{=}i)P(A{=}j)} \nonumber\\
&=\frac{n(P(\Xset,A)-P(\Xset)P(A))^2}{P(\Xset)P(\neg \Xset)P(A)P(\neg A)}=
\frac{n\delta^2(\Xset,A)}{P(\Xset)P(\neg \Xset)P(A)P(\neg A)}.
\end{align}
So, in principle, each term measures how much the observed frequency
$\fr(I_\Xset{=}i,A{=}j)$ deviates from its expectation $nP(I_\Xset{=}i)P(A{=}j)$ under the
independence assumption. If the data size $n$ is sufficiently large
and none of the expected frequencies is too small, the
$\chi^2$-measure follows approximately the $\chi^2$-distribution with one
degree of freedom. 
As a classical rule of thumb \citep{fisher1925}, the
$\chi^2$-measure can be used only, if all expected frequencies
$nP(I_\Xset{=}i)P(A{=}j)$, $i, j\in\{0,1\}$, are at least
5. However, the approximations can still be poor in some situations,
when the underlying binomial distributions are skewed, e.g., if $P(A)$
is near 0 or 1, or if $\fr(\Xset)$ and $\fr(\neg \Xset)$ are far from each other
\citep{yates,agresti}. According to \citet{carriere}, this is
quite typical for data in medical science. 

One reason for the inaccuracy of the $\chi^2$-measure is that the
original binomial distributions are discrete while the
$\chi^2$-distribution is continuous. A common solution is to make a {\em
  continuity correction} and subtract 0.5 from the expected frequency
$nP(\Xset)P(A)$. According to \citet{yates} the resulting continuity
corrected $\chi^2$-measure can give a good approximation to Fisher's
$p_F$, if the underlying hypergeometric distribution is not markedly
skewed. However, according to \citet{haber} the resulting
$\chi^2$-value can underestimate the significance, while the
uncorrected $\chi^2$-value overestimates it.

Mutual information is another popular asymptotic measure, which has
been used to test independence. For binary variables Eq.~\eqref{eqMIgeneral} becomes

\begin{equation}
\label{MImeasure}
\MI{=}\log \frac{P(\Xset A)^{P(\Xset A)}P(\Xset\neg A)^{P(\Xset\neg A)}P(\neg \Xset A)^{P(\neg \Xset A)}P(\neg \Xset\neg A)^{P(\neg \Xset\neg A)}}{P(\Xset)^{P(\Xset)}P(\neg \Xset)^{P(\neg \Xset)}P(A)^{P(A)}P(\neg A)^{P(\neg A)}}.
\end{equation}

Mutual information is actually an information theoretic measure, but
in statistics $2n\cdot \MI$ is known as log likelihood ratio or the $G$-test of independence. 
It follows asymptotically the $\chi^2$-dis\-tri\-bu\-tion \citep{wilks} and often it gives
similar results to the $\chi^2$-measure \citep{vilalta}. However, sometimes
the two tests can give totally different results \citep{agresti}.

\subsubsection*{Selecting the right model}

Selecting the right sampling model and defining the extremeness relation is a
controversial problem, which statisticians have argued for the last
century (see e.g., \citep{yates,agresti,lehmann1993,upton,howard}).
Therefore, we cannot give any definite recommendations which model to
select but each situation should be judged in its own context.

The main decision is whether the analysis should be done
conditionally or unconditionally and which variables $N$, $N_X$, or
$N_A$ should be considered fixed. In the multinomial model all
variables except $N=n$ are randomized. However, if the model is
conditioned with $N_X=\fr(\Xset)$, it leads to the double binomial
model. If the double binomial model is conditioned with
$N_A=\fr(A)$, it leads to the hypergeometric model. For
completeness, we could also consider the {\em Poisson model} where all
variables, including $N$, are unfixed Poisson variables. If the
Poisson model is conditioned with the given data size, $N=n$, it
leads to the multinomial model.\citep[ch. 4.6-4.7]{lehmann}

In principle, the sampling scheme should be decided before the data is
gathered. However, in pattern discovery the data may not be sampled
according to a particular scheme. In this situation the main choices
are to perform an unconditional analysis where none of the margins are
considered fixed or a conditional analysis where all margins are
considered fixed.  The main argument of the unconditional approach is
that the results are better generalizable outside the data set, if
some variables are kept unfixed. However, both multinomial and double
binomial models are computationally demanding, and in practice the
corresponding asymptotic tests have been used instead. The opponents
have argued that the unconditional approach is also conditional on the
data, since the unknown parameters ($p_X$ and/or $p_A$) are anyway
estimated from the observed counts ($\fr(\Xset)$ and/or
$\fr(A)$). Therefore, Fisher and his followers have suggested that we
should always assume both $N_X$ and $N_A$ fixed and use Fisher's exact
test or -- when it is heavy to compute -- a suitable asymptotic test. 

In pattern discovery the most popular choices for evaluating
dependency rules and other similar bipartition dependence patterns in
the variable-based interpretation have been Fisher's exact test (e.g.,
\citep{kingfkais,terada2013,terada2015,llinares2015,jabbar2016}) and the $\chi^2$-test (e.g.,
\citep{morishitasese,morishitanakaya,nijssen,fidep,causalrules,terada2015}). Both of these
tests have also been used for evaluating significance of improvement
(see Section \ref{sec:redsig}). According to our cross-validation
experiments \citep{kingfkais}, the $\chi^2$-measure can be quite
unreliable, in the sense that the discovered dependency rules may not
hold in the test data at all or their lift and leverage values differ
significantly between the training and test sets. The problem is
alleviated to some extent when the continuity correction is used, but
still the errors can be considerable. On the contrary, Fisher's $p$ has
turned out to be a very robust and reliable measure in the dependency
rule search and we recommend it as a first choice whenever applicable.
There is also an accurate approximation of Fisher's $p$ when faster
evaluation is needed \citep{fisherub}. Mutual information is also a
good alternative and it often produces the same rules as $p_F$.

\subsection{Sampling models for the value-based interpretation}
\label{subsec:valbased}

In the value-based interpretation the idea is that we would like to find
events $\Xset A$ or $\Xset \neg A$, which express a strong positive dependency,
even if the dependency between variables $I_\Xset$ and $A$ were 
relatively weak. In this case the strength of the dependency is
usually measured by lift, because leverage has the same
absolute value for all events $\Xset A$, $\Xset\neg A$, $\neg \Xset A$, 
$\neg \Xset\neg A$. However, lift alone is not a reliable 
measure, because it obtains its maximum value also when 
$\fr(\Xset A{=}i)=\fr(\Xset)=\fr(A{=}i)=1$ ($i\in \{0,1\}$) -- i.e., when the rule 
occurs on just one row \citep{hahsler}. Such a rule is quite likely due 
to chance and hardly interesting (see Example \ref{heartdiseaseexample}). 
Therefore, we should evaluate 
the probability of observing such a large lift
value, if $\Xset$ and $A$ were actually independent (independence
testing, $H_0$: $\Gamma=1$ \citep{benjaminileshno}) or, alternatively, that the
lift is at most some threshold $\gamma_0>1$ ($H_0$: $\Gamma\leq\gamma_0$ 
\citep{lallich}).

The $p$-value is defined like in the variable-based testing by
Eq.~\eqref{pvalue}. The only difference is how to define the 
extremeness relation $\ctable_i \succeq \ctable_j$. A necessary
condition for the extremeness of table $\ctable_i$ over $\ctable_j$ is that in
$\ctable_i$ the lift is larger than in $\ctable_j$. However, since the lift is
largest, when $N_X$ and/or $N_A$ are smallest (and $N_{X A}=N_X$ or
$N_{X A}=N_A$), it is sensible to require that also $N_{X A}$ is
larger in $\ctable_i$ than in $\ctable_j$.
If both $N_X$ and $N_A$ are fixed, then the lift is
larger than observed if and only if the leverage is larger than
observed, and it is enough to consider tables where $N_{X A}\geq
\fr(\Xset A)$. However, if either $N_X$, $N_A$, or both are unfixed, then we
should always check the lift $\Gamma=\frac{nN_{X A}}{N_X N_A}$
and compare it to the observed lift $\gamma(\Xset,A)$. 

In the following, we will describe different approaches for evaluating
statistical significance of dependency rules in the value-based
interpretation. The approaches fall into two categories depending on
whether the dependence is tested only in the part of data where the
rule antecedent holds or in the whole data. We will call these main
strategies {\em partial} and {\em complete evaluation of significance}
according to corresponding measures that are called {\em partial} and
{\em complete evaluators} \citep{vilalta}. We introduce three
approaches: partial evaluation with a single binomial test, complete
evaluation under the classical sampling models, and complete evaluation
with a single binomial test. Finally, we discuss the problem of
selecting the right model.

\subsubsection*{Partial evaluation with a single binomial test} 

In the previous research on association rules, some authors
\citep{dehaspe,lallich,lallich3,bruzzese,megiddosrikant} have
speculated how to test the null hypothesis $\Gamma=1$. For some 
reason, it has often been taken for granted that one should perform 
partial evaluation and evaluate significance of rule $\Xset \rightarrow A$ 
{\em in the part of the data where $\Xset$ is true}. As a solution, it has been 
suggested to use only a single binomial from the double binomial model.
This is equivalent to assuming two infinite baskets of apples, the red
and green one, but taking only a sample of $\fr(\Xset)$ apples from
the red basket and trying to decide whether there is a dependency
between the red colour and sweetness. It is assumed that 
$N_{X A}\sim Bin(\fr(\Xset),p_A)$ and the unknown parameter $p_A$ is 
estimated from the data, as usual. For positive dependence the $p$-value is 
defined as \citep{dehaspe}
\begin{equation}
\label{tradbin}
p=\sum_{i{=}\fr(\Xset A)}^{\fr(\Xset)}\left(\fr(\Xset) \atop i\right)P(A)^iP(\neg A)^{\fr(\Xset)-i}
\end{equation}
and for negative dependence as \citep{dehaspe}
\begin{equation}
p=\sum_{i{=}0}^{\fr(\Xset A)}\left(\fr(\Xset) \atop i\right)P(A)^iP(\neg A)^{\fr(\Xset)-i}.
\end{equation}

We see that $N_X{=}\fr(\Xset)$ is the only variable that has to be
fixed -- even $N$ can be unfixed. We note that since $N_A$ is unfixed,
$i$ goes from $\fr(\Xset A)$ to $\fr(\Xset)$ (and not to
$\min\{\fr(\Xset),\fr(A)\}$) in the case of positive dependence
\citep{lallich3}. The idea is that when $N_{X A}\geq \fr(\Xset A)$,
then $\frac{N_{X A}}{\fr(\Xset)}\geq P(A|\Xset)$, and since
$p_A{=}P(A)$ was fixed, then also $\Gamma\geq
\gamma(\Xset,A)$. Similarly, in the negative case $\Gamma\leq
\gamma(\Xset,A)$. So, the test checks correctly all cases where the
lift is at least as large (or as small) as observed.

Since the cumulative binomial probability is quite difficult to
calculate, it is common to estimate it asymptotically by the
$z$-score. 
The $z$-score measures how many standard deviations the
observed frequency deviates from its expectation.
In the case of positive dependence, the binomial variable
$N_{X A}$ has expected value $\avgsamp=\fr(\Xset)P(A)$ and standard
deviation $\sdsamp=\sqrt{\fr(\Xset)P(A)P(\neg A)}$. The corresponding
$z$-score is \citep{lallich3,bruzzese}

\begin{equation}
\label{z2measure}
z=\frac{\fr(\Xset A)-\avgsamp}{\sdsamp}=\frac{\fr(\Xset A)-\fr(\Xset)P(A)}{\sqrt{\fr(\Xset)P(A)P(\neg
A)}}
=\frac{\sqrt{n}\delta(\Xset,A)}{\sqrt{P(\Xset)P(A)P(\neg A)}}.
\end{equation}

If $\fr(\Xset)$ is sufficiently large and $P(A)$ is not too near to 1 or 0, the
$z$-score follows the standard normal distribution. However, when the
expected frequency $\fr(\Xset)P(A)$ is low (as a rule of thumb $<5$), the
binomial distribution is positively skewed. This means that the
$z$-score overestimates the significance. 

It is also possible to construct a partial evaluator from the mutual information
(Eq.~\eqref{MImeasure}) by ignoring terms related to $\neg \Xset$. The result is 
known as {\em $J$-measure} \citep{smythgoodman}:
\begin{equation}
\label{Jmeasure}
J=P(\Xset A)\log \frac{P(\Xset A)}{P(\Xset)P(A)}+P(\Xset\neg A)\log \frac{P(\Xset\neg
    A)}{P(\Xset)P(\neg A)}.
\end{equation}
However, it is an open problem how the corresponding $p$-value
could be evaluated and whether the $J$-measure could be used for
estimating statistical significance.

The problem of all partial evaluators is that two rules with different
antecedents $\Xset$ are not comparable. So, all rules (with different
$\Xset$) are thought to be from different populations and are tested
in different parts of the data.
We also note that the single binomial probability (like the double
binomial probability) is not an exchangeable measure in the sense that generally
$p(\Xset\rightarrow A)\neq p(A\rightarrow \Xset)$. The same holds for
the corresponding $z$-score and $J$-measure. This
can be counter-intuitive when the task is to search for statistical
dependencies, and these measures should be used with
care. In addition, with this binomial model the significance of the
positive dependence between $\Xset$ and $A$ is generally not the same
as the significance of the negative dependence between $\Xset$ and
$\neg A$. With the corresponding $z$-score the significance values
are related, and
$$\zpos(\Xset\rightarrow A)=-\zneg(\Xset\rightarrow \neg A),$$ where
$\zpos$ denotes the $z$-score of positive dependence and $\zneg$
the $z$-score of negative dependence. 
With the $J$-measure the significance of positive dependence between $\Xset$ and
$A$ and the significance of negative dependence between $\Xset$ and
$\neg A$ are equal.

\subsubsection*{Complete evaluation under the classical sampling models}

Let us now analyze the value-based significance of dependency rules
using the classical statistical models. For simplicity we consider
only positive dependence. We assume that the extremeness relation is
defined by lift $\Gamma$ and frequency $N_{X A}$, i.e., a contingency
table is more extreme than the observed contingency table, if
it has $\Gamma\geq \gamma(\Xset,A)$ and $N_{X A}\geq \fr(\Xset A)$.

In the multinomial model only the data size $N{=}n$ is fixed. Each
contingency table, described by triplet $\langle N_X, N_A, N_{X A}\rangle$, has 
probability $P(N_{X A},N_{X}-N_{X A},N_A-N_{X A},n-N_X-N_A+N_{X A}\mid n,p_X,p_A)$, defined by Eq.~\eqref{multinom}. 
The $p$-value is obtained when
we sum over all possible triplets where $\Gamma\geq \gamma(\Xset,A)$:
\begin{equation*}
p=\sum_{N_X{=}0}^n \sum_{N_{X A}{=}\fr(\Xset A)}^{N_X} \sum_{N_A{=}N_{X A}}^{Q_1}P(N_{X A},N_{X}-N_{X A},N_A-N_{X A},n-N_X-N_A+N_{X A}\mid n,p_X,p_A),
\end{equation*}
where $Q_1=\frac{nN_{X A}}{\gamma(\Xset,A) N_X}$. (We note that the terms are
zero, if $N_X<N_{X A}$.)

In the double binomial model $N_X=\fr(\Xset)$
is also fixed. Each contingency table, described by pair
$\langle N_A,N_{X A}\rangle$, has probability
$P(N_{X A},N_A-N_{X A}\mid n,\fr(\Xset),p_A)$ by Eq.~\eqref{doublebin}. 
Now we should sum over all possible pairs, where
$\Gamma\geq \gamma(\Xset,A)$:
\begin{equation*}
p=\sum_{N_{X A}{=}\fr(\Xset A)}^n \sum_{N_A{=}N_{X A}}^{Q_2} P(N_{X A},N_A-N_{X A}\mid n,\fr(\Xset),p_A),
\end{equation*}
where $Q_2=\frac{nN_{X A}}{\gamma(\Xset,A) \fr(\Xset)}$. 

In the hypergeometric model also $N_A=\fr(A)$ is fixed. As noted
before, the extremeness relation is now the same as in the
variable-based case and the $p$-value is defined by Eq.~\eqref{pfpos}.
This is an important observation, because it means that {\em Fisher's 
exact test tests significance also in the value-based 
interpretation}. The same is not true for the first two models,
where rule $\Xset \rightarrow A$ can get a different $p$-value in
variable-based and value-based interpretations.

\subsubsection*{Complete evaluation with a single binomial test} 

When $N_X$ and/or $N_A$ are unfixed the $p$-values are quite heavy to
compute. Therefore, we will now introduce a simple binomial model
(suggested in \citep{stataprkais} and as model 2 in \citep{lallich3}),
where it is enough to sum over just one variable. The binomial
probability can be further estimated by an equivalent $z$-score or the
$z$-score can be used as an asymptotic test measure as such. Contrary
to the previously described binomial test, this test performs a complete
evaluation in the whole data set, which means that the $p$-values of
different rules are comparable.

Let us suppose that we have an infinite basket of apples where the
probability of red and sweet apples is $p_{XA}$. According to
the independence assumption $p_{XA}=p_Xp_A$. A sample 
of $n$ apples is taken randomly from the basket. The probability of obtaining $N_{XA}$ sweet red apples among all $n$ apples is defined by binomial probability

\begin{equation*}
P(N_{X A}\mid n,p_X,p_A)=\left(n\atop N_{X A}\right) (p_Xp_A)^{N_{X A}}(1-p_Xp_A)^{n-N_{X A}}.
\end{equation*}

Since $N_{X A}$ is the only variable which occurs in the probability, the
extremeness relation is defined simply by $\ctable_i \succeq \obsctable
\Leftrightarrow N_{X A}\geq \fr(\Xset A)$. When the unknown parameters $p_X$ and $p_A$
are estimated from the data, the $p$-value of rule $\Xset\rightarrow A$ becomes 
\begin{equation}
\label{bintn}
  \pbin=\sum_{i{=}\fr(\Xset A)}^{n} \left({n \atop i}\right) (P(\Xset)P(A))^i (1-P(\Xset)P(A))^{n-i}.
\end{equation}

Since $N_{X A}$ is a binomial variable with expected value
$\avgsamp=nP(\Xset)P(A)$ and standard deviation
$\sdsamp=\sqrt{nP(\Xset)P(A)(1-P(\Xset)P(A))}$, the corresponding
$z$-score\index{$z$-score} is
\begin{equation}
\label{zscore}
  z(\Xset \rightarrow A)
  = \frac{\fr(\Xset A)-nP(\Xset)P(A)}{\sqrt{nP(\Xset)P(A)(1-P(\Xset)P(A))}}
  =\frac{\sqrt{n}\delta(\Xset,A)}{\sqrt{P(\Xset)P(A)(1-P(\Xset)P(A))}}.
\end{equation}
Because the discrete binomial distribution is approximated by the continuous normal 
distribution, the continuity correction can be useful, like with the
$\chi^2$-measure.

We note that this binomial probability and the corresponding $z$-score are
exchangeable, which is intuitively a desired
property. However, the statistical significance of positive dependence
between $\Xset$ and $A$ is generally not the same as the significance of
negative dependence between $\Xset$ and $\neg A$. For example, the
$z$-score for negative (or, equally, positive) dependence between $\Xset$ and
$\neg A$ is
$$z(\Xset\rightarrow \neg A)=\frac{\fr(\Xset\neg A)-nP(\Xset)P(\neg A)}{\sqrt{nP(\Xset)P(\neg A)(1-P(\Xset)P(\neg A))}}=\frac{-\sqrt{n}\delta(\Xset,A)}{\sqrt{P(\Xset)P(\neg A)(1-P(\Xset)P(\neg A))}}.$$

\subsubsection*{Selecting the right model}

The main decision in the value-based interpretation is whether the
significance of dependency rule $\Xset \rightarrow A$ is evaluated in
the whole data or only in the part of data where $\Xset$ holds. 
This decision is critical, because partial and complete evaluators can disagree 
significantly in their ranking and selection of rules. This is demonstrated in the 
following example.

\begin{table}[!h]
\begin{center}
\caption{Comparison of $p$-values and asymptotic measures for example 
rules $\Xset\rightarrow A$ and $\Yset\rightarrow A$.}
\label{patolesim}
\begin{tabular}{lll}
&$\Xset \rightarrow A$&$\Yset \rightarrow A$
{\rule{0pt}{2.2ex}}\\
\hline
$\pbinI$&1.06e-4&2.21e-5\\
$\pmul$&8.86e-13&1.01e-19\\
$p_F$&1.60e-12&7.47e-19\\
$\pdouble$&2.05e-13&7.35e-20\\
$z_1$&4.20&4.36\\
$\pbinII$&9.31e-10&8.08e-8\\
$z_2$&5.48&5.16\\
$J$&0.36&0.15\\
\end{tabular}
\end{center}
\end{table}

\begin{example}
\label{expathological}
Let us compare two rules, $\Xset\rightarrow A$ and $\Yset\rightarrow
A$, in the value-based interpretation. The frequencies are $n=100$,
$\fr(A)=50$, $\fr(\Xset)=\fr(\Xset A)=30$, $\fr(\Yset)=60$, and
$\fr(\Yset A)=50$, i.e., $P(A|\Xset)=1$ and $P(\Yset|A)=1$. The
$p$-values, $z$-scores, and $J$-values are given in Table
\ref{patolesim}.

All of the traditional association rule measures (binomial $\pbinII$,
Eq.\ \eqref{tradbin}, its $z$-score $z_2$, and $J$-measure) favour
rule $\Xset\rightarrow A$, while all the other measures (binomial
$\pbinI$, Eq.\ \eqref{bintn}, its $z$-score $z_1$, multinomial
$\pmul$, double binomial $\pdouble$, and Fisher's $p_F$) rank rule
$\Yset\rightarrow A$ better. In the three classical models, the
difference between the rules is quite remarkable.
\end{example}

In general, we do not recommend partial evaluation for dependency rule mining. 
The main problem is that the $p$-values of discovered
rules are not comparable, because each of them has been tested in a
different part of data. In addition, the measures are not exchangeable,
which means that $\Xset\rightarrow A$ can get a totally different
ranking than $A\rightarrow \Xset$, even if they express the same
dependency between events.

When the classical statistical models are used, the only difference to
the variable-based interpretation is that now the discrepancy measure
is lift. Computationally, the only practical choices are the
hypergeometric model and asymptotic measures. The hypergeometric model
produces reliable results, but it tends to favour large leverage
instead of lift, which might be more interesting in the value-based
interpretation. In addition, one should check for each rule
$\Xset\rightarrow A$ that the dependency is due to strong
$\gamma(\Xset,A)$ and not due to $\gamma(\neg\Xset,\neg A)$. With this
checking the $\chi^2$-measure can also be used. According to our
experiments \cite[Ch.~5]{whthesis}, the $\chi^2$-measure and the
$z$-score (Eq.~\eqref{zscore}) tend to find rules with the
strongest lift (among all compared measures), but at the same time the
results are also the most unreliable. Robustness of the
$\chi^2$-measure can be improved with the continuity correction, but
with the $z$-score it has only a marginal effect. One solution is to
use the $z$-score only for preliminary pruning and select the rules
with the corresponding binomial probability 
\citep{stataprkais}. Based on these considerations, we cannot give 
a universal recommendation, but Fisher's exact test is always a safe
choice, if there is no specific need to maximize lift. If large lift
values are desired, one could consider either the $\chi^2$-measure or
the $z$-score accompanied by an exact binomial test.

\section{Redundancy and significance of improvement}
\label{sec:redsig}

An important task in dependency rule discovery is to identify
redundant rules, which add little or no additional information on
statistical dependencies to other rules.  In this section we consider
an important type of redundancy called {\em superfluousness}, where a more
specific dependency rule does not improve its generalization
significantly. We present statistical significance tests for
evaluating superfluousness in the value-based and variable-based
interpretations. Finally, we will briefly discuss the relationship to
a more general approach of speciousness testing.

\subsection{Redundant and superfluous rules}
\label{subsec:redoverview}

According to a classical definition \citep{bastide} ``An association rule is redundant if 
it conveys the same information – or less general information – than the information
conveyed by another rule of the same usefulness and the same relevance.'' However, 
what is considered useful or relevant depends on the modelling purpose, 
and numerous definitions of redundant or uninformative rules have been proposed.

In traditional association rule research, the goal has been to
find all sufficiently frequent and `confident' (high precision)
rules. Thus, if the sufficient frequency or precision of a rule can 
be derived from other rules, the rule can be considered
redundant (e.g.,
\citep{aggarwal2001,goethalsderivablerules,ChengKe2008,LiHamilton2004};
see also a good overview by \cite{balcazarred2010}). On the other hand, when the goal is
to find statistical dependency rules, then rules that are merely
side-products of other dependencies can be considered
uninformative. An important type of such 
dependencies are superfluous specializations ($\Xset \rightarrow
A$) of more general dependency rules ($\Yset \rightarrow A$, $\Yset
\subsetneq \Xset$). This concept of superfluous rules covers earlier
notions of non-optimal or superfluous classification rules
\citep{jiuyongli}, (statistically) redundant rules
\citep{hurao,kingfkais} and unproductive rules
\citep{webbml}. 

Superfluous rules are a common problem, because rules `inherit'
dependencies from their ancestor rules unless their extra factors
reverse the dependency. This is regrettable, because undetected
superfluous rules may lead to quite serious misconceptions.  For
example, if disease $D$ is caused by gene group $\Yset$ (i.e., $\Yset
\rightarrow D$), we are likely to find a large number of other
dependency rules $\Yset\Qset \rightarrow D$ where $\Qset$ contains
coincidental genes.  Now one could make a conclusion that the
combination $\Yset\Qset_1$ (with some arbitrary $\Qset_1$) predisposes
to disease $D$ and begin preventive care only with these patients.

Intuitively, the idea of superfluousness is clear. A superfluous rule
$\Xset\rightarrow A$ contains extraneous variables
$\Qset\subsetneq \Xset$ which have no effect or only weaken the
original dependency $\Xset\setminus\Qset \rightarrow A$. It is also
possible that $\Qset$ apparently improves the dependency but the
improvement is spurious (due to chance). 
In this case the apparent improvement occurs only in the sample, and it may be 
detected with appropriate statistical significance
tests. We recall that significance tests do not necessarily detect all
superfluous rules but we can
always adjust the significance level to prune more or less potentially
superfluous rules. Formalizing the idea of superfluousness is more difficult,
because it depends on the used measure, assumed statistical model,
required significance level, and -- most of all -- whether we are
using the value-based or variable-based interpretation. Therefore, we
give here only a tentative, generic definition of superfluousness. 

\begin{definition}[Superfluous dependency rules]
\label{def:redundancy}
Let $\measure$ be a goodness measure which is used to evaluate dependency rules. 
Let us assume that $\measure$ is increasing by goodness 
and rule $\Xset \rightarrow A{=}i$ improves $\Yset\rightarrow A{=}i$, $i\in\{0,1\}$, when $\measure(\Xset \rightarrow A{=}i)>\measure(\Yset\rightarrow A{=}i)$
(for decreasing measures, $\measure(\Xset \rightarrow A{=}i)<\measure(\Yset\rightarrow A{=}i)$). 
Let $\Model$ be a statistical model which is used for determining the statistical significance and $\alpha$ the selected significance level. 

Rule $\Xset \rightarrow A{=}i$ is superfluous (given $\measure$, $\Model$, and $\alpha$) 
if there exists rule $\Yset \rightarrow A{=}i$, $\Yset \subsetneq \Xset$, such that 
\begin{itemize}
\item[(i)] $\measure(\Xset \rightarrow A{=}i)\leq\measure(\Yset\rightarrow A{=}i)$ 
(vs.~$\measure(\Xset \rightarrow A{=}i)\geq\measure(\Yset\rightarrow A{=}i)$)
or 
\item[(ii)] Improvement of rule $\Xset \rightarrow A{=}i$ over rule $\Yset\rightarrow A{=}i$ is not significant at level $\alpha$ (value-based interpretation) or 
\item[(iii)] Improvement of rule $\Xset \rightarrow A{=}i$ over rule $\Yset\rightarrow A{=}i$ 
is less significant than the improvement of rule $\neg \Xset\rightarrow A{\neq}i$ over rule $\neg Y\rightarrow A{\neq}i$ (variable-based interpretation).
\end{itemize}
\end{definition}

We note that in a special case where $P(\Xset)=P(\Yset)$, rules 
$\Xset \rightarrow A{=}i$ and $\Yset\rightarrow A{=}i$, $\Yset\subsetneq \Xset$, 
have equivalent contingency tables and they obtain the same measure value with all commonly used goodness measures (that are functions of $N_X$, $N_A$, $N_{XA}$ and $n$). Otherwise, if $P(\Xset)<P(\Yset)$, the contingency tables are different and rule $\Xset \rightarrow A{=}i$ may or may not improve $\Yset\rightarrow A{=}i$ depending on the observed counts and the selected goodness measure. 
The special case $\Yset \subsetneq \Xset$, $P(\Xset)=P(\Yset)$, is closely connected to the notions of {\em closed itemsets} ($\Xset$ such that $\forall \Zset\supsetneq \Xset$: $P(\Xset)>P(\Zset)$) and their {\em minimal generators} ($\Yset\subseteq \Xset$ such that $P(\Yset)=P(\Xset)$ and $\nexists \Yprimeset\subsetneq \Yset$: $P(\Yprimeset)=P(\Yset)$) \citep{pasquier99,bastide}. If the rule antecedents $\Xset$ are selected only among closed sets, some of them may have distinct minimal generators $\Yset\subsetneq \Xset$ and are necessarily superfluous. This is avoided, if the rule antecedents are selected only among minimal generators (also called {\em free sets} \citep{boulicaut00}), but the rules may still be superfluous when tested against more general rules.

\subsection{Testing superfluousness in the value-based interpretation}
\label{subsec:redvariablebased}

Let us first consider the problem of superfluousness in the
value-based interpretation, where the significance tests are somewhat
simpler.  To simplify notations, we will consider only rule
$\Xset\rightarrow A$ with a positive-valued consequent. For
$\Xset\rightarrow \neg A$ the tests are analogous, except $A$ is
replaced by $\neg A$. 

In traditional association rule research, the goodness measure
$\measure$ is precision (or, equivalently, lift, because the
antecedent is fixed).  Rule $\Xset \rightarrow A$ is called {\em
  productive}, if $P(A|\Xset)>P(A|\Yset)$ for all $\Yset\subsetneq \Xset$ 
(e.g., \citep{bayardogunopulos,webbml}). The significance of
productivity is tested separately for all $\Yset\rightarrow A$,
$\Yset\subsetneq \Xset$, and all $p$-values should be below some fixed
threshold $\alpha$. 

Let us now notate $\Xset=\Yset\Qset$
(i.e., $\Qset=\Xset\setminus \Yset$, $\Qset \neq \emptyset$) so that we can compare rule $\Yset\Qset\rightarrow A$ to a simpler rule $\Yset\rightarrow A$.
In each test, the null hypothesis is that there is no improvement in the precision: 
$P(A|\Yset\Qset)=P(A|\Yset)$. The condition means that $\Qset$
and $A$ are conditionally independent given $\Yset$. The
significance is estimated by calculating $p(\Qset\rightarrow A\mid\Yset)$,
i.e., the $p$-value of rule $\Qset \rightarrow A$ in the set where $\Yset$
holds. Now it is quite natural to assume $\fr(\Yset)$, $\fr(\Yset\Qset)$, and $\fr(\Yset A)$ fixed, which leads to the hypergeometric model. The corresponding test is Fisher's exact test for conditional independence, and the significance
of productivity of $\Yset\Qset \rightarrow A$ over $\Yset\rightarrow A$ is
\citep{webbml}
\begin{equation}
\label{eq:signofprod}
p(\Qset\rightarrow A\mid\Yset)=\sum_{j{=}0}^{J_1} \frac{\left(\fr(\Yset\Qset) \atop \fr(\Yset\Qset A)+j\right) \left(\fr(\Yset\neg \Qset) \atop \fr(\Yset\neg \Qset A)-j\right)}{\left(\fr(\Yset) \atop \fr(\Yset A) \right)},
\end{equation}
where $J_1=\min\{\fr(\Yset\Qset \neg A),\fr(\Yset\neg\Qset A)\}$.
When the $\chi^2$-measure is used to estimate the significance of
productivity, the equation is \citep{liuhsuma}
\begin{equation}
\chi^2(\Qset \rightarrow A\mid\Yset)=\frac{\fr(\Yset)(P(\Yset)P(\Yset\Qset A)-P(\Yset\Qset)P(\Yset A))^2}{P(\Yset\Qset)P(\Yset\neg \Qset)P(\Yset A)P(\Yset \neg A)}.
\end{equation}

In principle, measure $\measure$ can be any goodness measure for statistical
dependence between values, including the previously introduced
binomial probabilities and corresponding $z$-scores. However,
different measures can disagree on their ranking of rules and which
rules are considered superfluous. For example, leverage has a strong
bias in favour of general rules, when compared to lift or precision. This 
is clearly seen from expression 
$\delta(\Yset,A)=P(\Yset)(P(A|\Yset)-P(A))=P(\Yset)(\gamma(\Yset,A)-1)$.
On the other hand, asymptotic measures like the $z$-score and the
$\chi^2$-measure tend to
overestimate the significance, when the frequencies are
small. Therefore, it is possible that a rule is not superfluous,
when evaluated with an asymptotic measure, but superfluous, when the
exact $p$-values are calculated. 

\subsection{Testing superfluousness in the variable-based interpretation}
\label{subsec:redvaluebased}

In the variable-based interpretation, superfluousness of dependency rules
is more difficult to judge, because there may be two kinds of
improvement into opposite directions in the same time. Improvement of
rule $\Yset\Qset\rightarrow A$ over rule $\Yset\rightarrow A$ is 
tested as in the value-based interpretation. 
However, in the same time rule $\neg \Yset\rightarrow \neg A$ may improve a more general rule 
$\neg(\Yset\Qset)\rightarrow \neg A$, and one should weigh which 
improvement is more significant. 

The significance of improvement of rule $\neg \Yset\rightarrow \neg A$ over  
$\neg(\Yset\Qset)\rightarrow \neg A$ is tested in the same way as productivity of 
$\Yset\Qset\rightarrow A$ over $\Yset\rightarrow A$.
However, now the null hypothesis is conditional independence between $\neg\Yset$ and $\neg A$ given $\neg(\Yset\Qset){=}\neg\Yset\vee\Yset\neg\Qset$. It is natural to assume $\fr(\neg(\Yset\Qset))$, $\fr(\neg \Yset)$, and $\fr(\neg(\Yset\Qset)\neg A)$ fixed, which leads to an exact test

\begin{equation}
\label{eq:signofnegatedprod}
p(\neg \Yset\rightarrow \neg A\mid\neg(\Yset\Qset))=\sum_{j{=}0}^{J_2} \frac{\left(\fr(\neg \Yset) \atop \fr(\neg\Yset \neg A)+j\right) \left(\fr(\Yset\neg \Qset) \atop \fr(\Yset\neg \Qset \neg A)-j\right)}{\left(\fr(\neg(\Yset\Qset)) \atop \fr(\neg(\Yset\Qset) \neg A) \right)},
\end{equation}

where $J_2=\min\{\fr(\neg\Yset A),\fr(\Yset\neg\Qset \neg A)\}$.
The corresponding $\chi^2$-test is
\begin{equation}
\chi^2(\neg \Yset\rightarrow \neg A\mid \neg(\Yset\Qset))=
\scalebox{1.25}{$\frac{\fr(\neg(\Yset\Qset))(P(\neg(\Yset\Qset))P(\neg\Yset \neg A)-P(\neg \Yset)P(\neg(\Yset\Qset)\neg A))^2}{P(\neg\Yset)P(\Yset\neg \Qset)P(\neg(\Yset\Qset)A)P(\neg(\Yset\Qset)\neg A)}$}.
\end{equation}

An important property of variable-based superfluousness testing is that sometimes significance tests can be avoided altogether. This is possible with such 
goodness measures, for which any improvement is significant
improvement. One such measure is $p_0$, the first and largest term of $p_F$.
It can be shown \citep{HamalainenWebb17} that 
for dependency rules $\Yset\rightarrow A$ and $\Yset\Qset\rightarrow A$ holds
$$\frac{p_0(\Yset\Qset\rightarrow A)}{p_0(\Yset\rightarrow A)}=\frac{p_0(\Qset\rightarrow A\mid\Yset)}{p_0(\neg\Yset\rightarrow \neg A\mid\neg(\Yset\Qset))}.$$
It is an open problem whether the equality
holds exactly for the cumulative probability, $p_F$, but at least it  holds 
approximately. 
This is also the justification for the simpler
superfluousness testing in Kingfisher \citep{kingfkais}, where a dependency rule is
considered superfluous if it has a larger (poorer) $p_F$-value than some
of its ancestor rules.

Previously, we have already shown that goodness measures for
the variable-based and value-based interpretations can diverge quite much
in their ranking of rules. The same holds for superfluousness testing. The
following example demonstrates that the same rule 
may or may not be superfluous depending on the interpretation.

\begin{example}
Let us reconsider the rules $\Xset\rightarrow A$ ($=\Yset\Qset \rightarrow A$) and 
$\Yset\rightarrow A$ in Example \ref{expathological}. Rule $\Xset \rightarrow A$ is 
clearly productive with respect to $\Yset \rightarrow A$ ($P(A|\Yset)=1.00$ vs.\ $P(A|\Xset)=0.83$). Similarly, rule $\neg \Yset\rightarrow \neg A$ is productive with respect to $\neg \Xset\rightarrow \neg A$ ($P(\neg A|\neg \Yset)=1.00$ vs.\ $P(\neg A|\neg \Xset)=0.71$).

Let us now calculate the significance of productivity using Fisher's exact test. In the value-based interpretation, we evaluate only the first improvement:
$$p(\Qset\rightarrow A\mid\Yset)=\frac{\left(\fr(\Yset\neg \Qset) \atop \fr(\Yset\neg \Qset\neg A)\right)}{\left(\fr(\Yset) \atop \fr(\Yset A)\right)}=\frac{\left(30 \atop 10\right)}{\left(60 \atop 50\right)}=3.99\cdot 10^{-4}.$$
The value is so small that we can assume that the productivity is
significant and $\Xset\rightarrow A$ is not superfluous.

In the variable-based interpretation, we evaluate also the second improvement:
$$p(\Yset\rightarrow \neg A\mid \neg (\Yset\Qset))=\frac{\left(\fr(\Yset\neg \Qset) \atop \fr(\Yset\neg \Qset\neg A)\right)}{\left(\fr(\neg(\Yset\Qset)) \atop \fr(\neg(\Yset\Qset) \neg A)\right)}=\frac{\left(30 \atop 10\right)}{\left(70 \atop 50\right)}=1.86\cdot 10^{-10}.$$
This value is much smaller than the previous one, which means that the improvement of $\neg\Yset \rightarrow \neg A$ over $\neg\Xset \rightarrow \neg A$ is more significant than the improvement of $\Xset\rightarrow A$ over $\Yset\rightarrow A$. Thus, we would consider rule $\Xset\rightarrow A$ superfluous. We would have ended up into the same conclusion, if we had simply compared the $p_F$-values of both rules: $p_F(\Yset\rightarrow A)=7.47\cdot 10^{-19}<1.60\cdot 10^{-12}=p_F(\Xset\rightarrow A)$.
\end{example}

\subsection{Relationship to speciousness}
\label{subsec:redspeciousness}

The concept of superfluousness is closely related to  
{\em speciousness} \citep{yule1903,HamalainenWebb17}, where
an observed unconditional dependency vanishes or changes its sign when
conditioned on other variables, called {\em confounding factors}. The
latter phenomenon, reversal of the direction of the dependency, is
also known as {\em Yule-Simpson's paradox}.  In the context of
dependency rules, rule $\Xset \rightarrow A$ is considered specious if
there is another rule $\Yset \rightarrow A$ or $\Yset \rightarrow \neg
A$ such that $\Xset$ and $A$ are either independent or negatively
dependent {\em in the population} when conditioned on $\Yset$ and
$\neg\Yset$. In the sample either of the conditional dependencies may
also appear as weakly positive, and one has to test their significance
with a suitable test, like Birch's exact test \citep{birch},
conditional mutual information \citep{HamalainenWebb17} or various
$\chi^2$-based tests.

It is noteworthy that the confounding factor $\Yset$ does not necessarily
share any attributes with $\Xset$. However, in a special case when
$\Yset\subsetneq \Xset$, Birch's exact test for speciousness of
$\Xset\rightarrow A$ with respect to $\Yset \rightarrow A$ reduces to
Eq.~\eqref{eq:signofprod} (significance of productivity). On the
other hand, Birch's exact test for speciousness of $\Yset\rightarrow
A$ with respect to $\Xset\rightarrow A$ is equivalent to 
Eq.~\eqref{eq:signofnegatedprod}. So, testing superfluousness of $\Xset
\rightarrow A$ with respect to $\Yset \rightarrow A$ in a
variable-based interpretation can be considered as a special case of
testing if $\Xset\rightarrow A$ is specious by $\Yset \rightarrow A$
or vice versa.

\section{Dependency sets}
\label{sec:sets}

Dependency rules capture the most common conception of dependence as a
relationship between two elements. Often, however, multiple elements
will all interact with each other, and the mutual dependency structure
is better represented by set-type of patterns. Dependency sets are a
general name for set-type patterns that express interdependence
between the elements of the set. In this section we will first give a
short overview of set dependency patterns and then describe key
approaches for evaluating their statistical significance.

\subsection{Overview}
\label{subsec:setsoverview}

Approaches to finding dependency sets differ in terms of the forms of
interdependence that they seek to capture. A common starting point is
to assume mutual dependence among the elements of the set,
i.e.,\ absence of mutual independence (Definition
\ref{defmutindep}). However, this notion is very inclusive because it
suffices that $\Xset$ contains at least one subset $\Yset\subseteq
\Xset$ where $P(\Yset)\neq \prod_{A_i\in \Yset}P(A_i)$. This means
that the property of mutual dependence is monotonic, i.e., all
supersets of a mutually dependent set are also mutually dependent.  To
avoid an excessive number of patterns, dependency sets usually
represent only some of all mutually dependent sets, like minimal
mutually dependent sets \citep{brinmotwani}, sets that present new
dependencies in comparison to their subsets (for some $A\in \Xset$,
$\delta(\Xset\setminus \{A\},A)\neq 0$) \citep{meo}, or sets for which
all bipartitions express statistical dependence (for all
$\Yset\subsetneq \Xset$ $\delta(\Yset,\Xset\setminus \Yset)\neq 0$)
\citep{Webb10}. We note that the latter two approaches assume 
bipartition dependence (absence of bipartition independence, 
Definition \ref{defbipindep}), which is a stronger condition than mutual 
dependence. 

Compared to dependency rules, dependency sets offer a more compact
presentation of dependencies, and in some contexts the reduction in
the number of patterns can be quite drastic. This is evident when we
recall that any set $\Xset$ can give rise up to $|\Xset|$ rules of the form
$\Xset\setminus\{A_i\}\rightarrow A_i$ and up to $2^{|\Xset|}-2$ rules of the form $\Xset\setminus \Yset\rightarrow \Yset$. In many cases these permutation
rules reflect the same statistical dependency. This is always true
when $|\Xset|=2$ ($A\rightarrow B$ and $B\rightarrow A$ present the same
dependency), but the same phenomenon can occur also with more complex
sets as the following observation demonstrates.
\begin{observation}
Let $\Xset$ be a set of binary attributes such that 
for all $\Yset\subsetneq \Xset$ $P(\Yset)=\prod_{A_i\in \Yset}P(A_i)$
(i.e., attributes are mutually independent). Then 
for all $\Zset\subsetneq \Xset$ $\delta(\Xset\setminus \Zset,\Zset)=P(\Xset)-\prod_{A_i\in \Xset\setminus \Zset}P(A_i)\prod_{A_i\in \Zset}P(A_i)=P(\Xset)-\prod_{A_i\in \Xset}P(A_i)$.
\end{observation}
This means that when all proper subsets of $\Xset$ express only mutual
independence, then all permutation rules of $\Xset\setminus \Zset
\rightarrow \Zset$ have the same leverage, frequency and expected
frequency, and many goodness measures would rank them equally good.
In real world data, the condition holds seldom precisely, but the same
phenomenon tends to occur to some extent also when all subsets express
at most weak dependence. In this case, it is intuitive to report only
set $\Xset$ instead of listing all of its permutation rules.

In principle, all dependency rules could be represented by dependency
sets without losing any other information than the division to an
antecedent and a consequent. The reason is that for any dependency
rule $\Xset\setminus \Yset \rightarrow \Yset$, set $\Xset$ is mutually
dependent. This follows immediately from the fact that mutual
independence of $\Xset$ (Definition \ref{defmutindep}) implies
bipartition independence between $\Xset\setminus \Yset$ and $\Yset$
(Definition \ref{defbipindep}) for any $\Yset\subsetneq \Xset$.
However, as explained below, some set dependency approaches have more
stringent constraints which may exclude interesting dependency rules
selected under other schemes. Further, if mutual independence is
violated only by a single bipartition, or if the objective is to find
dependencies with a specific element of interest, a dependency rule
between the relevant partitions will more concisely convey the
relevant information. Which dependency rule or set scheme is most
appropriate depends entirely on the analytic objective.

The approaches for finding statistically significant dependency sets
can be roughly divided into two categories: i) selecting dependency
sets among frequent item sets and testing their statistical
significance afterwards and ii) searching directly all sufficiently
strong and significant dependency sets using appropriate goodness
measures and significance tests.  In the following subsections we
describe the main methods for evaluating statistical significance in
these approaches.

\subsection{Statistically significant dependency sets derived from candidate frequent itemsets}
\label{subsec:setsfreq}

Frequent itemsets \citep{agrawal96} are undoubtedly the most popular
type of set patterns in knowledge discovery. A frequent itemset is a
set of true-valued binary attributes (called items, according to the
original market-basket setting) whose frequency exceeds some
user-specified minimum frequency threshold ('minimum
support'). However, being frequent does not ensure that the elements
in an itemset express statistical dependence. For example, consider
two elements $A$ and $B$ that each occur in all but one example such
that the examples in which $A$ and $B$ do not occur differ. In this
case itemset $\{A,B\}$ will occur in all but two examples and thus be
frequent, but it will represent negative dependence rather than
positive dependence such as association discovery typically seeks.

Frequent itemsets have been employed as an initial step in dependency
set discovery in order to constrain the number of patterns that must
be considered. The idea is to search first all frequent itemsets and then 
select statistically significant dependency set patterns among them.
A limitation is that this approach will fail to discover statistically significant but infrequent 
dependency sets, which can be the most significant.

The most common null hypothesis used for significance testing of
dependency sets is mutual independence between all attributes of the
data (Definition \ref{defmutindep}). Statistical significance of a
set is defined as the probability that its frequency is at least as
large as observed, given the mutual independence assumption.  In
principle, any significance testing approach could be used, but often
this is done with randomization testing. In swap randomization (e.g.,
\citep{gionismannila,swaprand}), both column margins (attribute
frequencies) and row margins (numbers of items on each row) are kept
fixed. The latter requirement allows suppression of dependencies that
are due to co-occurrence of items only due to their appearing solely
in rows that contain many items. A variant is iterative randomization
\citep{hanhijarvi2009tell}. This approach begins with fixed row and
column margins, but on each iteration it adds the most significant
frequent itemset as a new constraint. The randomization problem is
computationally very hard, and thus it is sufficient that the
frequencies of itemsets hold only approximately. The process is
repeated until no more significant itemsets can be found.

\cite{Vreeken2014} have proposed identifying statistically significant
dependency sets $\Xset$ using the binomial test for the null
hypothesis of independence of events $A_i\in\Xset$ 
(Eq.~\ref{weakercondition}). Under the independence assumption, the probability 
of $\Xset$ in the population is $p_X=\prod_{A_i\in\Xset}p_{A_i}$, where 
$p_{A_i}$s can be estimated by observed $P(A_i)$s as usual. Then the probability of observing $N_X\geq \fr(\Xset)$ in a sample of $n$ rows is 
\begin{equation}
\pbin=\sum_{j=\fr(\Xset)}^n \left(n \atop
j\right)\left[\prod_{A_i\in \Xset} P(A_i)\right]^j\left[1-\prod_{A_i\in \Xset}
P(A_i)\right]^{n-j}.
\end{equation}
We note that this test assumes a weaker notion of independence than mutual 
independence (Definition \ref{defmutindep}). In consequence, it may find 
fewer dependency patterns than the previously described randomization test 
for mutual independence.

\subsection{Direct search for significant dependency sets}
\label{subsec:setsdirect}

An alternative approach is to search directly for sets that satisfy
specific criteria of statistical dependence and significance, using
those criteria to prune the search space and support efficient search.
Sometimes, these set patterns are still called `rules' or are
represented by the best rule that can be derived from the set.
Examples are correlation rules \citep{brinmotwani}, strictly
non-redundant association rules \citep{stataprkais,fidep}, and -- the
most rigorous of all -- self-sufficient itemsets
\citep{Webb10,WebbVreeken13}. All these pattern types have three
common requirements: dependency set $\Xset$ expresses mutual
dependence, it adds new dependencies to its subsets $\Yset\subsetneq
\Xset$, and the dependency is significant with the selected measure.

Correlation rules \citep{brinmotwani} are defined as minimal sets $\Xset$, where $\Xset$ expresses mutual dependence (at least for some $\xbf\in \Dom(\Xset)$ $P(\Xset=\xbf)$ is greater than expected under independence) but all $\Yset\subsetneq \Xset$ express mutual independence. The significance is evaluated with the $\chi^2$-measure
\begin{equation}\chi^2=\sum_{(a_1,\dots,a_m)\in\{0,1\}^m}\frac{n\left(P(\Xset=(a_1,\dots,a_m))-\prod_{A_i\in\Xset}P(A_i{=}a_i)\right)^2}{\prod_{A_i\in\Xset}P(A_i{=}a_i)}
\end{equation}
with one degree of freedom. Since all supersets of $\Xset$ can only
increase the $\chi^2$-value, only minimal sets whose $\chi^2$-value
exceeds a specified threshold are presented.

Strictly non-redundant association rules are an intermediate form
between set type and rule type patterns, where each set is presented
by its best rule, whose significance is evaluated in the desired
sampling model. The discovered patterns are mutually dependent sets
$\Xset$ that express bipartition dependence between some $A\in \Xset$
and $\Xset\setminus \{A\}$ and the bipartition dependence is more
significant than any bipartition dependence in simpler sets
$\Yset\subsetneq \Xset$ (between any $B\in\Yset$ and $\Yset \setminus
\{B\}$). In the significance testing, one can assume either value- or
variable-based interpretation and use any of the sampling models
presented in Section \ref{sec:deprules}. For search purposes, feasible
choices are the binomial model and the corresponding $z$-score
\citep{stataprkais} for the value-based interpretation, the double
binomial test and the corresponding $\chi^2$-measure \citep{fidep} for
the variable-based interpretation, and Fisher's exact test that can be
used in both interpretations.

Self-sufficient itemsets are a pattern type that imposes much stronger
requirements. The core idea is that an itemset should only be
considered interesting if its frequency cannot be explained by
assuming independence between any partition of the items. That is,
there should be no partition $\Qset\subsetneq\Xset,
\Xset\setminus\Qset$ such that $P(\Xset)\approx
P(\Qset)P(\Xset\backslash\Qset)$. For example, being male ($M$) and
having prostate cancer ($P$) are associated and hence should form a
dependency set $\{M,P\}$. Suppose that having a name containing a `G'
($G$) is independent of both factors. Then $\{M,P,G\}$ should not be
a dependency set. However, it is more frequent than would be expected
by assuming independence between $\{M\}$ and $\{P,G\}$ or between
$\{M,G\}$ and $\{P\}$, and hence most interestingness measures would
assess both $\{P,G\}\rightarrow\{M\}$ and $\{M,G\}\rightarrow \{P\}$
as interesting. Nonetheless, under the self-sufficient itemset
approach $\{M,P,G\}$ can be discarded because it is not more frequent
than would be expected by assuming independence between $\{G\}$ and
$\{M,P\}$.

In self-sufficient itemsets this requirement is formalized as a test
for \emph{productivity}. It is required that there is a significant
positive dependency between every partition of the itemset, when
evaluated with Fisher's exact test. In addition, self-sufficient
itemsets have two additional criteria: they have to be non-redundant
and independently productive.

In the context of self-sufficient itemsets, set $\Xset$ is considered redundant, if
\begin{equation}
\exists_{\Yset\subsetneq \Xset,\Zset\subsetneq \Yset}\fr(\Yset) = \fr(\Zset).
\end{equation}
The motivation is that if $A$ is a necessary consequent of another
set of items $\Zset$, then $\Yset=\{A\} \cup \Zset$ should be
associated with everything with which $\Zset$ is associated.  For
example $\Zset = \{pregnant\}$ entails $A=\mathit{female}$ ($\Yset =
\{female, pregnant\}$) and \emph{pregnant} is associated with
\emph{oedema}. In consequence, $\Xset = \{female, pregnant, oedema\}$
is not likely to be interesting if $\{\mathit{pregnant},
\mathit{oedema}\}$ is known.

A final form of test that can be employed is whether the frequency of
an itemset $\Xset$ can be explained by the frequency of its productive
and nonredundant supersets $\Yset\supsetneq\Xset$. For example, if $A,
B$ and $C$ are jointly necessary and sufficient for $D$ then all
subsets of $\{A, B, C,D\}$ that include $D$ should be productive and
nonredundant. However, they may be misleading, as they fail to capture
the full conditions necessary for $D$. \cite{webbcritical} proposes
that if $\Yset\supsetneq\Xset$ is productive and nonredundant, $\Xset$
should only be considered potentially interesting if it is
\emph{independently productive}, meaning that it passes tests for
productivity when data covered by $\Yset\setminus\Xset$ are not
considered.

\section{Multiple testing problem}
\label{sec:multtest}

The goal of pattern discovery is to find all sufficiently good
patterns among exponentially many possible candidates. This leads
inexorably to the problem of multiple hypothesis testing. The core of
this problem is that as the number of tested patterns increases, it
becomes ever more likely that spurious patterns pass their tests,
causing type I error.
	
In this section, we will first describe the main principles and
popular correction methods that the statistical community has
developed to remedy the problem. Such understanding is critical to
addressing this issue in the pattern discovery context. We then
introduce some special techniques for increasing the power to detect
true patterns while controlling the number of false discoveries in the
pattern discovery context.

\subsection{Overview}
\label{subsec:multtestoverview}

The problem of multiple testing is easiest to demonstrate in the 
classical Neyman-Pearsonian hypothesis testing. Let us suppose we are
testing $m$ true null hypotheses (spurious patterns) and in each test the 
probability of type I error
is exactly the selected significance level $\alpha$. (In general, the
probability is at most $\alpha$, but with increasing power it
approaches $\alpha$.) In this case the expectation is that in every $m\cdot \alpha$
tests a type I error is committed and a spurious pattern passes the
test.  With normal significance levels this can be quite a
considerable number. For example if $\alpha=0.05$ and 100 000 spurious patterns 
are tested, we can expect 5000 of them to pass the test.

\begin{figure}[!h]
\begin{center}
\includegraphics[width=0.65\textwidth]{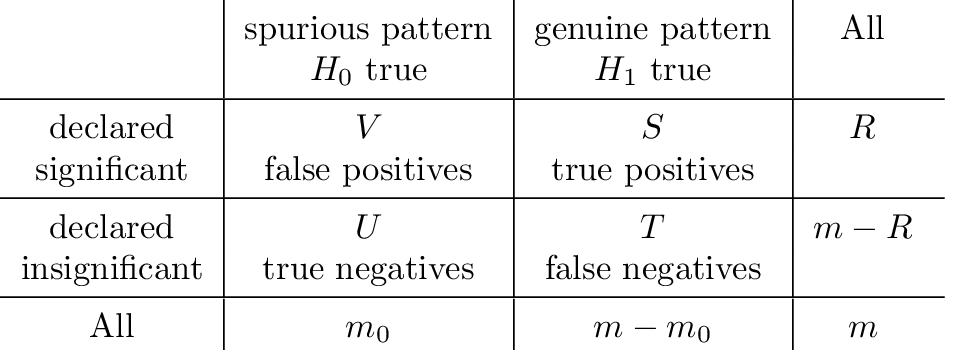}
\caption{A contingency table for $m$ significance tests. Here $m_0$ is an unknown parameter for the number of true null hypotheses and $R$ is an observable random variable for the number of rejected hypotheses. $S$, $T$, $U$, and $V$ are unobservable random variables.}
\label{table:signtests}
\end{center}
\end{figure}

Solutions to the multiple testing problem try to control type I errors
among all tests. In practice, there are two main approaches: The
traditional approach is to control the {\em familywise error rate} which
is the probability of accepting at least one false discovery (rejecting
a true null hypothesis). Using the notations of Fig.~\ref{table:signtests}  
the familywise error rate is $FWER=P(V\geq 1)$. Another, less stringent
approach is to control the {\em false discovery rate}, which is the
expected proportion of false discoveries,
$FDR=E\left(\frac{V}{\max\{R,1\}}\right)=E\left(\frac{V}{R}\mid R>0\right)P(R>0)$. 

Since $FDR\leq FWER$, control of $FWER$ subsumes control of $FDR$. In
a special case where all null hypotheses are true ($m=m_0$),
$FWER=FDR$. The latter means that a $FDR$ controlling method controls
$FWER$ in a {\em weak sense}, when the probability of type I errors is
evaluated under the {\em global null hypothesis} $H_0^C=\cap_{i=1}^m
H_i$ (all $m$ hypotheses are true). However, usually it is required
that $FWER$ should be controlled in a {\em strong sense}, under any
set of true null hypotheses. (For details, see e.g.,
\citep{gedudoit}.)

In general, $FWER$ control is preferred when false discoveries are
intolerable (e.g., accepting a new medical treatment) or when it is
expected that most null hypotheses would be true, while $FDR$ control
is often preferred in exploratory research, where the number of
potential patterns is large and false discoveries are less serious
(e.g., \citep{goeman2011}).
 
\subsection{Methods for multiple hypothesis testing} 
\label{subsec:multtestmethods}

The general idea of multiple hypothesis testing methods is to make
rejection of individual hypotheses more difficult by adjusting the
significance level $\alpha$ (or the corresponding critical value of
some test statistic) or, equivalently, adjusting individual
$p$-values, $p_1,\hdots, p_m$, corresponding to hypotheses
$H_1,\hdots,H_m$. When the goal is to control $FWER$ at level
$\alpha$, the procedure determines for each hypothesis $H_i$ an
adjusted significance level $\hat{\alpha}_i$ (possibly
$\hat{\alpha}_1=\hdots=\hat{\alpha}_m$) such that $H_i$ is rejected if
and only if $p_i\leq \hat{\alpha}_i$. Alternatively, the $p$-value of
$H_i$ can be adjusted and the adjusted $p$-value $\hat{p}_i$ is
compared to the original significance level $\alpha$.  Now $FWER$ can
be expressed as
\begin{equation}
\label{eqfwer}
FWER=P(\cup_{i\in K} \{P_i\leq \hat{\alpha}_i\})
=P(\cup_{i\in K} \{\hat{P}_i\leq \alpha\}),
\end{equation}
where $K$ is the set of indices of true null hypotheses and $P_i$ and $\hat{P}_i$ denote 
random variables for the original and adjusted $p$-values.

The correction procedures are designed such that $FWER\leq \alpha$
holds at least asymptotically, when the underlying assumptions are
met.  In addition, it is usually required that the adjusted $p$-values
have the same order as the original $p$-values, i.e., $p_i\leq p_j
\Leftrightarrow \hat{p}_i\leq \hat{p}_j$. This `monotonicity of
$p$-values' is by no means necessary for $FWER$ control, but it is in
line with the statistical intuition according to which a pattern
should not be declared significant if a more significant pattern (with
smaller $p$) is declared spurious
\cite[p. 65]{westfallyoung}. Reporting adjusted $p$-values (together
with original unadjusted $p$-values) is often recommended, since they
are more informative than binary rejection decisions. However,
individual $p$-values cannot be interpreted separately because
$\hat{p}_i$ is the smallest $\alpha$-level of the entire test
procedure that rejects $H_i$ given $p$-values or tests statistics of
all hypotheses \citep{gedudoit}.

Famous multiple testing procedures and their assumptions are listed in Table 
\ref{MHTmethods}. Bonferroni and  
\v{S}\'{i}d\'{a}k corrections as well as the single-step $minP$ method are examples of {\em single-step methods}, where the same adjusted significance level is applied to all hypotheses. All the other methods in the table are {\em step-wise methods} that determine individual significance levels for each hypothesis, depending on the order of $p$-values and rejection of other hypotheses. Step-wise methods can be further divided into {\em step-down methods} (Holm-Bonferroni method and the step-down $minP$ method by \cite{westfallyoung}) that process hypotheses in the ascending order by their $p$-values and {\em step-up methods} (Hochberg, Benjamini-Hochberg and Benjamini-Hochberg-Yekutieli methods) that proceed in the opposite order. In general, single-step methods are least powerful and step-up methods most powerful, with the exception of the powerful $minP$ methods.

Powerful methods are always preferable since they can detect most
true patterns, but the selection of the method depends on also other factors like 
availability of all $p$-values during evaluation, the expected
proportion of true patterns (false null hypotheses), seriousness of false 
discoveries, and assumptions on the dependency structure between hypotheses. 
In the following we will briefly discuss these issues. For more details, we refer the 
interested reader to e.g., \citep{goeman2014,gedudoit}.

The least powerful method for controlling $FWER$ is the popular
Bonferroni correction.  The lack of power is due to two pessimistic
assumptions: $m_0$ is estimated by its upperbound $m$ and the
probability of type I error by upperbound $P((P_1\leq
\frac{\alpha}{m})\vee \hdots \vee (P_m\leq \frac{\alpha}{m}))\leq
\sum_{i=1}^mP(P_i\leq \frac{\alpha}{m})$ (Boole's or Bonferroni's
inequality). Therefore, the Bonferroni correction is least powerful
when many null hypotheses are false or the hypotheses are positively
associated. The \v{S}\'{i}d\'{a}k correction \citep{sidak} is slightly
more powerful, because it assumes independence of true null hypotheses
and can thus use a lower upperbound for the probability of type I
error.  However, the method gives exact control of $FWER$ only under
the independence assumption. The control is not guaranteed if the true
hypotheses are negatively dependent and the method may be overly
conservative if they are positively dependent. The Bonferroni and
\v{S}\'{i}d\'{a}k corrections are attractive for pattern discovery
where the size of the space of alternatives can be predetermined,
because they impose minimal computational burden requiring simply that
the value of $\alpha$ be appropriately decreased.

The Holm-Bonferroni method \citep{holm-bonferroni} is a sequential
variant of the Bonferroni method. It proceeds in a step-wise manner,
by comparing the smallest $p$-value to $\frac{\alpha}{m}$ like the
Bonferroni method, but the largest to $\alpha$. Therefore, it rejects
always at least as many null hypotheses as the Bonferroni method and
the gain in power is greatest when most null hypotheses are false. The
Hochberg method \citep{hochberg} can be considered as a step-up
variant of the Holm-Bonferroni method. It is a more powerful method
especially if there are many false null hypotheses or the $p$-values
of false null hypotheses are positively associated. However, it has
extra requirements for the dependency structure among
hypotheses. Sufficient conditions for the Hochberg method (and the
underlying Simes inequality) are independence and certain types of
positive dependence (e.g., {\em positive regression dependence on a
  subset} \citep{benjaminiyekutieli}) between true
hypotheses. \v{S}\'{i}d\'{a}k's method can also be implemented in a
similar step-wise manner by using criterion
$p_i>1-(1-\alpha)^{1/(m-i+1)}$ in the Holm-Bonferroni method. However,
the resulting Holm-\v{S}\'{i}d\'{a}k method assumes also independence
of hypotheses.

The Benjamini-Hochberg method \citep{benjamini} and
the Ben\-jamini-Hochberg-Yekutieli method \citep{benjaminiyekutieli} are
also step-up methods, but they control $FDR$ instead of
$FWER$. The Benjamini-Hochberg method is always at least as powerful as
the Hochberg method, and the difference is most pronounced when there are
many false null hypotheses. The Benjamini-Hochberg method is also based on the Simes inequality and has 
the same requirements for the dependency structure between true hypotheses 
(independence or certain types of positive dependence). 
The Benjamini-Hochberg-Yekutieli method allows
also negative dependencies, but it is less powerful and may sometimes
be even more conservative than the Holm-Bonferroni method
\citep{goeman2014}.

These step-wise approaches are problematic in pattern discovery unless
statistical testing is applied as a postprocessing step.  This is
because they require all null hypotheses to be sorted on $p$-value
which implies that all $p$-values must be known before the corrections
are applied. However, step-wise methods are easily applied in
multi-stage procedures (e.g., \citep{webbml,komiyama2017statistical})
that first select constrained sets of candidate patterns which are
then subsequently subjected to statistical testing.

\begin{table}[!ht]
\begin{center}
\caption{Multiple hypothesis testing procedures for controlling $FWER$ or $FDR$ at level $\alpha$ and assumptions on the dependency structures between true $H_i$s. The methods are presented in a uniform manner leading to the rejection of $H_1,\hdots,H_{k}$ (keeping $H_{k+1},\hdots,H_m$). It is assumed that the hypotheses are ordered by their $p$-values and $p_1\leq \hdots \leq p_k \leq \hdots p_m$. $P_j$ is a random variable for the $p$-value of hypothesis $H_j$.}
\label{MHTmethods}
\begin{tabular}{llll}
{\bf Method}&{\bf Control}&{\bf Criterion}&{\bf Assumptions on $H_i$s} {\rule{0pt}{2.6ex}}\\
\hline
Bonferroni&$FWER$&$k=\min(i): p_{i+1}>\frac{\alpha}{m}$&none {\rule{0pt}{2.6ex}}\\
\v{S}\'{i}d\'{a}k&$FWER$&$k=\min(i): p_{i+1}>1-(1-\alpha)^{\frac{1}{m}}$&independence (or positive\\
&&&dependence)\\
Holm-Bonferroni&$FWER$&$k=\min(i): p_{i+1}>\frac{\alpha}{m-i}$&none\\
Hochberg&$FWER$&$k=\max(i): p_{i}\leq \frac{\alpha}{m-i+1}$&independence or certain\\
&&&types of positive dependence\\ 
single-step $minP$&$FWER$&$k=\min(i): P(\min\limits_{1\leq j\leq m} P_j\leq p_i)>\alpha$&none\\
step-down $minP$&$FWER$&$k=\min(i): P(\min\limits_{i\leq j\leq m} P_j\leq p_i)>\alpha$&none\\
Benjamini-Hochberg&$FDR$&$k=\max(i): p_k\leq \frac{k\cdot \alpha}{m}$&independence or certain\\
&&&types of positive dependence\\
Benjamini-Hochberg-&$FDR$&$k=\max(i): p_k\leq \frac{k\cdot \alpha}{m\cdot c(m)}$, where&none\\
Yekutieli&&$c(m)=\sum_{i=1}^m \frac{1}{i}\approx \ln(m)+0.58$&\\
&&if negative dependence and&\\
&&$c(m)=1$ otherwise&\\
\end{tabular}
\end{center}
\end{table}

The $minP$ methods \citep{westfallyoung} present a different
approach to multiple hypothesis testing. These methods are usually
implemented with permutation testing or other resampling methods, and
thus they adapt to any dependency structure between null
hypotheses. This makes them powerful methods and they have been shown
to be asymptotically optimal for a broad class of testing problems
\citep{meinshausen}. 

The $minP$ methods are based on an alternative expression of $FWER$ (Eq.~\eqref{eqfwer}): $FWER=P(\cup_{i\in K} \{P_i\leq \hat{\alpha}\}\mid H_K)=P(\min_{i\in K} \{P_i\leq \hat{\alpha}\}\mid H_K)$,
where $H_K$ is an intersection of all true hypotheses and $\hat{\alpha}$ is an adjusted significance level. Therefore, an optimal $\hat{\alpha}$ can be determined as an $\alpha$-quantile from the distribution of the minimum $p$-value among the true null hypotheses, i.e.,
$$\hat{\alpha}=\max \{a\mid P(\min P_i\leq a\mid H_K)\leq \alpha\}.$$
In principle, any technique for estimating the $\alpha$-quantile can be used, but analytical methods are seldom available. However, the evaluation can be done also empirically, with resampling methods.

For strong control of $FWER$ the probability should be evaluated under
$H_K$, which is unknown. Therefore, the estimation is done under the
complete null hypotheses $H_0^C$. Strong control (at least {\em
  partial strong control} \citep{rempala2013}) can still be obtained
under certain extra conditions. One such condition is {\em subset
  pivotality} \citep[p. 42]{westfallyoung} that requires the raw
$p$-values (or other test statistics) of true null hypotheses to have
the same joint distribution under $H_0^C$ and any other set of
hypotheses. Since the true null hypotheses are unknown, the minimum
$p$-value is determined among all null hypotheses (the single-step
method) or among all unrejected null hypotheses (the step-down
method). The resulting single-step adjustment is
$\hat{p}_i=P(\min\limits_{1\leq j\leq m}P_j\leq p_i\mid H_0^C)$. A
similar adjustment can be done with other test statistics $\measure$,
if subset pivotality or other required conditions hold. Assuming that
high $\measure$-values are more significant, the adjusted $p$-value is
$\hat{p}_i=P(\max\limits_{1\leq j\leq m}\measure_j\leq t_i\mid
H_0^C)$, where $\measure_i$ is a random variable for the test
statistic of $H_i$ and $\measureval_i$ is its observed value. The
probability under $H_0^C$ can be estimated with permutation testing,
by permuting the data under $H_0^C$ and calculating the proportion of
permuted data sets where $\min P$ or $\max \measure$ value is at least
as extreme as the observed $p_i$ or $t_i$. In pattern discovery
complete permutation testing is often infeasible, but there are more
efficient approaches combining the $minP$ correction with approximate
permutation testing (e.g.,
\citep{hanhijarvi2011,minato2014,llinares2015}. However, the time and
space requirements may still be too large for many practical pattern
mining purposes.

\subsection{Increasing power in pattern discovery}
\label{subsec:multtestpower}

In pattern discovery the main problem of multiple hypothesis testing
is the huge number of possible hypotheses. This number is the same as
the number of all possible patterns or the size of the search space
that is usually exponential. If correction is done with respect to all
possible patterns, the adjusted critical values may become so small
that few patterns can be declared as significant. This means that one
should always use as powerful correction methods as possible or
control $FDR$ instead of $FWER$ when applicable, but this may still be
insufficient. A complementary strategy is to reduce the number of
hypotheses or otherwise target more power to those hypotheses that are
likely to be interesting or significant. In the following we describe
three general techniques designed for this purpose: hold-out
evaluation, filtering hypotheses and weighted hypothesis testing.

The idea of {\em hold-out evaluation} \citep{webbml} (also known as
{\em two-stage testing}, e.g., \citep{miller2001}) is to use only a
part of the data for pattern discovery and save the rest for testing
significance of patterns. The method consists of three steps:

\begin{itemize}
\item[(i)] Divide the data into an exploratory set $\data_E$ and a hold-out set $\data_H$.
\item[(ii)] Search for patterns in $\data_E$ and select $k$ patterns for testing. Note that 
the selection process at this step can use any principle suited to the application and 
need not involve hypothesis testing.
\item[(iii)] Test the significance of the $k$ patterns in $\data_H$ using any multiple 
hypothesis testing procedure. 
\end{itemize}

Now the number of hypotheses is only $k$ which is typically much less
than the size of the search space. This makes the method powerful,
even if the $p$-values in the hold-out set are likely larger than they
would have been in the whole data set. The power can be further
enhanced by selecting powerful multiple testing methods, including
methods that control $FDR$.  A potential problem of hold-out
evaluation is that the results may depend on how the data is
partitioned. In a pathological case a pattern may occur only in the
exploratory set or only in the hold-out set and thus remain
undiscovered \citep{liuzhang2011}.

Another approach is to use {\em filtering methods} (see e.g.,
\citep{bourgon2010}) to select only promising hypotheses for
significance testing. Ideally, the filter should prune out only true
null hypotheses without compromising control of false discoveries. In
practice, the true nulls are unknown and the filter uses some data
characteristics to detect low power hypotheses that are unlikely to
become rejected. Unfortunately, some filtering methods affect also the
distribution of the test statistic and can violate strong control of $FWER$. 
As a solution it has been suggested that the
filtering statistic and the actual test statistic should be
independent given a true null hypothesis \citep{bourgon2010}.

In pattern discovery one useful filtering method is to prune out so
called {\em untestable hypotheses} \citep{teradaLAMP2013,mantel1980}
that cannot cause type $I$ errors. This approach can be used when
hypothesis testing is done conditionally on some data characteristics,
like marginal frequencies. The idea is to determine a priori, using
only the given conditions, if a hypothesis $H_i$ can ever achieve
sufficiently small $p$-value to become rejected at the adjusted level
$\hat{\alpha}_i$. If this is not possible (i.e., if the smallest
possible $p$-value, $p_i^*$, would be too large,
$p_i^*>\hat{\alpha}_i$), then the hypothesis is called
'untestable'. Untestable hypotheses cannot contribute to $FWER$ and
they can be safely ignored when determining corrections for multiple
hypothesis testing.

For example, the lowest possible $p$-value with Fisher's exact test
for rule $A\rightarrow B$ with any contingency table having marginal
frequencies $\fr(A)=10$, $\fr(B)=4$ and $n=20$ is $p^*=0.043$.  If we
test just one hypothesis with $\alpha=0.05$, then this pattern can
pass the test and the hypothesis is considered testable. However, if
we test two hypotheses and use Bonferroni correction, then the
corrected $\alpha$-level is $\hat{\alpha}=0.025$ and the hypothesis is
considered untestable ($p^*=0.043>0.025=\hat{\alpha}$).

In practice, selecting testable hypotheses can improve the power of the method
considerably.  In pattern discovery the idea of testability has been utilized
successfully in the search algorithms, including the LAMP procedure 
\citep{teradaLAMP2013,minato2014} that controls $FWER$ with the Bonferroni
correction and Westfall-Young light \citep{llinares2015} that implements a 
$minP$ method.

A third approach is to use a {\em weighted multiple testing procedure}
(e.g., \citep{finos2007,holm-bonferroni}) that gives more power to
those hypotheses that are likely to be most interesting.  Usually, the
weights are given a priori according to assumed importance of
hypotheses, but it is also possible to determine optimal weights from
the data to maximize power of the test (see e.g.,
\citep{roeder2009}). The simplest approach is an {\em allocated
  Bonferroni procedure} \citep{weightedbonferroni} that allocates
total $\alpha$ among all $m$ hypotheses according to their
importance. Each hypothesis $H_i$ is determined an individual
significance level $\hat{\alpha}_i$ such that $\sum_{i=1}^m
\hat{\alpha}_i\leq \alpha$. This is equivalent to a weighted
Bonferroni procedure, where one determines weights $w_i$ such that
$\sum_{i=1}^m w_i=m$ and reject $H_i$ if $p_i\leq
\frac{w_i\alpha}{m}$. There are also weighted variants of other
multiple correction procedures like the weighted Holm-Bonferroni
procedure \citep{holm-bonferroni} and the weighted Benjamini-Hochberg
procedure \citep{weightedBH}. Usually, these procedures do not respect
the monotonicity of $p$-values, which means that the most significant
patterns may be missed if they were deemed uninteresting.

In the pattern discovery context one natural principle is to base the
weighting on the complexity of patterns and favour simple patterns.
This approach is used in the {\em method of layered critical values}
\citep{webbcritical} where simpler patterns are tested with looser
thresholds and strictest thresholds are reserved to most complex
patterns. The motivation is that simpler patterns tend to have higher
proportions of significant patterns and can be expected to be more
interesting. In addition, this weighting strategy supports efficient
search, because it helps to prune deeper levels of the search space.

When dependency patterns are searched, the complexity can be
characterized by the number of attributes in the pattern which is the
same as the level of the search space.  \citet{webbcritical} has
suggested an allocation strategy where all patterns at level $L$ are
tested with threshold $\hat{\alpha}_L$ such that
$\sum_{L=1}^{L_{\max}}\hat{\alpha}_L \cdot \mathit{S_L}\leq \alpha$,
where $L_{\max}$ is the maximum level and $\mathit{S_L}$ is the number
of all possible patterns at level $L$. One such allocation is to
set $$\hat{\alpha}_L=\frac{\alpha}{L_{\max}\mathit{S_L}}.$$ The method
was originally proposed for the breadth-first search of classification
rules, but it can be applied to other pattern types and depth-first
search as well. The only critical requirement is that the bias towards
simple patterns fits the research problem. In a pathological case, the
method may miss the most significant patterns if they are too
complex. However, the same patterns might remain undetected also with
a weaker but more balanced testing procedure.

\section{Conclusions}
\label{sec:concl}

Pattern discovery is a fundamental form of exploratory data analysis.
In this tutorial we have covered the key theory and techniques for
finding statistically significant dependency patterns that are likely
to represent true dependencies in the underlying population.
We have concentrated on two general classes of patterns: dependency rules
that express statistical dependencies between condition and consequent
parts and dependency sets that express mutual dependence between
set elements.

Techniques for finding true statistical dependencies are based on
statistical tests of different types of independence. The general idea
is to evaluate how likely it is that the observed or a stronger
dependency pattern would have occurred in the given sample data, if
the independence assumption had been true. If this probability is too
large, the pattern can be discarded as having a high risk of being
spurious.  In this tutorial we have presented the core relevant
statistical theory and specific statistical tests for different
notions of dependence under various assumptions on the underlying
sampling model.

However, in many applications it is often desirable to use stronger
filters to the discovered patterns than a simple test for independence.
Statistically significant dependency rules and sets can be generated
by adding unrelated or even negatively associated elements to existing
patterns.  Unless further tests are also satisfied, such as tests for
productivity and significant improvement, the discovered rules and
sets are likely to be dominated by many superfluous or redundant
patterns. Fortunately, statistical significance testing can also be
employed to control the risk of `discovering' these and other forms of
superfluous patterns. We have also surveyed the key such techniques.

The final major issue that we have covered is that of multiple
testing. Each statistical hypothesis test controls the risk that its
null hypothesis would be rejected if that hypothesis were false.
However, typical pattern discovery tasks explore exceptionally large
numbers of potential hypotheses, and even if the risks for each of the
individual hypotheses are extremely small, they can accumulate until
the cumulative risk of false discoveries approaches almost certainty.
We also survey multiple testing methods that can control this
cumulative risk.

The field of statistically sound pattern discovery is in its infancy
and there are numerous open problems. Most work in the field has been
restricted to attribute-value or transactional data. Patterns over
more complex data types like sequences and graphs would benefit also
from statistically sound techniques but may require new statistical
tests to be feasible. The possibilities of allowing for untestable
hypotheses are also opening many possibilities for substantially
increasing the power of multiple testing procedures. The field has
been dominated by frequentist approaches of significance testing, but
there is much scope for application of Bayesian techniques. But
perhaps the two biggest challenges are determining the right
statistical tests to identify patterns of interest for specific
applications and then developing efficient search algorithms that find
the most significant patterns under those tests.

It is important to remember that statistical significance testing
controls only the risk of false discoveries -- type I error. It does
not control the risk of type II error -- of failing to discover a
pattern. When sample sizes are reasonably large, it is reasonable to
expect that statistically sound pattern discovery techniques will find
all real strong patterns in the data and will not find spurious weak
patterns. However, it is important to recognize that in some
circumstances it will be more appropriate to explore alternative
techniques that trade off the risks of type I and type II error.

Statistically sound pattern discovery has brought the field of pattern
mining to a new level of maturity, providing powerful and robust
methods for finding useful sets of key patterns from sample data. We
hope that this tutorial will help bring the power of these techniques
to a wider group of users.


\appendix
\section*{Appendix: Terminology}
\label{app:terminology}

The following lists some similar looking terms that have a
different meaning in the traditional pattern discovery and statistics.\\

{\bf Association} (Statistics)\\ 
Generally, statistical dependence between two (or more) random
variables. In a narrower sense, it refers to statistical dependence
between categorical variables, while the word 'correlation' is used
for numerical variables.

{\bf Association rule} (Pattern discovery)\\ 
Traditionally, an association rule $\Xset\rightarrow \Yset$ merely
means sufficiently frequent co-occurrence of two attribute sets,
$\Xset$ and $\Yset$. The sufficient frequency is defined by a
user-specified threshold $min_{\fr}$. In the traditional definition
\citep{agrawalass}, it has also been required that the 'association'
should be sufficiently strong, measured by precision
('confidence'). As such, an association rule does not necessarily
express any statistical dependence or the dependence may be negative,
instead of the assumed positive dependence. Therefore, it has become
more common to require that the rule expresses positive statistical
dependence, i.e.  $P(\Xset \Yset)>P(\Xset)P(\Yset)$, and use
statistical dependence measures like lift or leverage instead of
precision. In this paper, we call rule-formed statistical dependencies
without any necessary minimum frequency requirements as 'dependency
rules'.

{\bf Confidence} (Pattern discovery)\\ 
A traditional measure for the strength of an association rule $\Xset
\rightarrow \Yset$ defined as $\phi(\Xset\rightarrow
\Yset)=P(\Yset|\Xset)$. In pattern recognition, information retrieval,
and binary classification, this measure is called 'precision' or
'positive predictive value'.

{\bf Confidence interval and confidence level} (Statistics)\\ 
Confidence interval is an interval estimate of some unknown population
parameter. Confidence level (e.g., 95\%) determines the proportion of
confidence intervals that contain the true value of the parameter. The
concepts are closely related to statistical hypothesis testing: a
confidence interval with confidence level $1-\alpha$ contains all
values $s_0$ for which the corresponding null hypothesis $S=s_0$ is
not rejected at significance level $\alpha$.

{\bf Support} (Pattern discovery)\\ 
A term used for frequency in frequent itemset and association rule
mining. The support of an attribute set (itemset) $\Xset$ can mean
either absolute frequency, $\fr(\Xset)$, or relative frequency,
$P(\Xset)$. The support of association rule $\Xset\rightarrow \Yset$
usually means $\fr(\Xset \Yset)$ or $P(\Xset \Yset)$ but sometimes it
can refer to $\fr(\Xset)$ or $P(\Xset)$. The latter are also called
'coverage' of the rule, although coverage can sometimes refer to
$\fr(\Xset\Yset)$ or $P(\Xset\Yset)$.

{\bf Support} (Mathematics, Statistics)\\ 
In general, the support of a function is the set of points where the
function is not zero-valued or the closure of that set. In the
probability theory and statistics, the support of a distribution whose
density function is $f$, is the smallest closed set $S$ such that
$f(x)=0$ for all $x \notin S$.

\bibliographystyle{spbasic}      


\end{document}